\documentclass[]{aastex631}
\usepackage{array}
	\shorttitle{Phase mixing and the 1/\textit{f} spectrum}
	\shortauthors{Magyar \& Van Doorsselaere}
	\graphicspath{{./}{figures/}}
\begin{document}
			
			\title{Phase mixing and the 1/\textit{f} spectrum in the solar wind }

			\correspondingauthor{Norbert Magyar}
			\email{norbert.magyar@kuleuven.be}
			
			\author[0000-0001-5731-8173]{Norbert Magyar} \thanks{FWO postdoctoral fellow}
			\affiliation{Centre for mathematical Plasma Astrophysics (CmPA), Department of Mathematics, 
				KU Leuven, Celestijnenlaan 200B bus 2400, \\ B-3001 Leuven, Belgium}

			\author[0000-0001-9628-4113]{Tom Van Doorsselaere}
			\affiliation{Centre for mathematical Plasma Astrophysics (CmPA), Department of Mathematics,
				KU Leuven, Celestijnenlaan 200B bus 2400, \\ B-3001 Leuven, Belgium}
			
			
			
			\begin{abstract}
				The origin and evolution of the 1/\textit{f} power law observed in the energy spectrum of solar coronal and solar wind fluctuations at scales of around an hour is not entirely understood. Several existing theories aim at explaining it, involving both linear and nonlinear mechanisms. An often overlooked property of the solar corona and solar wind is their highly inhomogeneous nature. In this paper we investigate the linear evolution of pure Alfv\'en and surface Alfv\'en waves propagating through a plasma which is inhomogeneous across the magnetic field. The inhomogeneity is given by density, which we model to be two-dimensional colored noise, with power spectral slopes ranging from -2 to -1. Alfv\'en waves propagate independently on individual magnetic field lines, and eventually get completely out of phase through the process of phase mixing, leading to unrealistic spectra. When the coupling between the inhomogeneous background and the propagating waves is fully accounted for, transverse waves such as surface Alfv\'en waves (also referred to as kink or Alfv\'enic) appear, showing collective wave behavior of neighboring magnetic field lines with different Alfv\'en speeds. We show that the linear cascade of surface Alfv\'en wave energy, induced by phase mixing and resonant absorption, leads to a perpendicular wave energy spectrum which tends to the perpendicular power spectrum of the background density. Based on our model, we propose that a perpendicular density power spectrum of 1/\textit{f} in the solar corona can induce, through linear processes, the 1/\textit{f} spectrum of the fluctuations that is observed at the largest scales. 
				
			\end{abstract}
			
		\keywords{MHD Waves, Solar Wind}
		
		
		\section{Introduction} \label{sec:intro}
		
		In-situ measurements of the solar wind reveal velocity and magnetic field fluctuations of a turbulent nature, displaying a well-defined power-law spectrum across orders of magnitude in frequency, part of which is identifiable with the inertial range of the turbulent cascade.   
		At the largest scales, the fluctuations usually display a 1/\textit{f} scaling, also known as `pink noise', in what is referred to as the energy-containing scale \citep{1982JGR....87.3617B,1982JGR....87.2215D}. The 1/\textit{f} range is most prominent in the fast wind, but it is also present in the slow wind \citep{2019A&A...627A..96B}, and was shown to exist from 0.3 AU out to 4.8 AU \citep{2009EM&P..104..101B} and even inside the solar corona \citep{2021PhRvL.127y5101K}. Interestingly, the frequency power spectrum of propagating Alfv\'enic waves seen by CoMP below 1.3 $\mathrm{R_\odot}$ also display a 1/$f$ scaling, more likely associated with the parallel wavenumber of these waves \citep{2015NatCo...6E7813M}. A spectral break at the higher end of the 1/\textit{f} scale (in frequency) occurs around $10^{-3}$ Hz in the fast wind at 1 AU, and it shifts to higher (lower) frequencies with decreasing (increasing) distance from the Sun \citep{2013LRSP...10....2B}. The shifting of the spectral break is an important clue regarding the nature of the 1/\textit{f} range. One explanation is that turbulence develops at larger and larger scales with increasing distance from the Sun, `eating away' the energy contained in the 1/\textit{f} range and establishing an inertial range instead \citep[e.g.,][]{1986PhRvL..57..495M}. Nevertheless, the real nature and origin of the 1/\textit{f} range are still not clear, and numerous explanations were put forth to date. One of the earliest explanations is that the 1/\textit{f} range is the remnant of scale-invariant processes occurring near the solar surface \citep{1986PhRvL..57..495M,2007ApJ...657L.121M}. Details of the nonlinear cascade of Alfv\'en waves, such as nonlocal interactions and inverse cascade \citep{2007PhRvE..76c6305D,2011PhRvE..83f6318D} or the coherence of Els\"{a}sser variables as a result of continuous reflection \citep{1989PhRvL..63.1807V}, combined with linear effects related to reflection at the transition region \citep{2012ApJ...750L..33V} were shown to lead to a 1/\textit{f} spectrum at large scales. Inspired by the effect of strong shears on hydrodynamical turbulence, \citet{2016Ap&SS.361..364G} suggested that shears in the solar wind can have the same effect on the existing turbulence, and could lead to a similar 1/\textit{f} spectrum at the largest scales. It was also shown that the parametric decay and subsequent inverse cascade of Alfv\'en waves propagating through the solar corona and wind can lead to a 1/\textit{f} spectrum \citep{2018JPlPh..84a9006C}. Yet another recent theory highlights that properties of Alfv\'en waves in the solar wind, such as their spherical polarization \citep[e.g.,][]{1971JGR....76.3534B,2004AnGeo..22.3751B}, naturally induce a cutoff magnitude for perturbations, at double the background magnetic field magnitude, which translates to the 1/\textit{f} scale observed in the spectrum at large scales \citep{2018ApJ...869L..32M}. \par 
		The diversity of the existing explanations suggests that an 1/\textit{f} range can arise in the solar wind from many different circumstances and effects, and it is not clear whether the 1/\textit{f} range is the result of linear or nonlinear processes, or both. There is evidence that at low frequencies (inside the 1/\textit{f} range) the magnetic field spectrum evolves following the linear WKB model, while at higher frequencies (inside the inertial range) the evolution is faster, indicative of turbulent dynamics \citep{1982JGR....87.3617B,1990JGR....9511945M}. Some of the predictions of above presented theories could soon be tested by measurements from the Parker Solar Probe, aiding in narrowing down the set of possible explanations. There is already some preliminary evidence that predictions involving parametric decay \citep{2018JPlPh..84a9006C} are consistent with the appearance of the 1/\textit{f} range inside the solar corona \citep{2021PhRvL.127y5101K}. Additionally, the 1/$f$ spectrum of waves observed by CoMP below 1.3 $R_\odot$ \citep{2015NatCo...6E7813M} might indicate that this range is set by the wave driver at the solar surface, or otherwise develops during wave propagation already over short distances. Despite the existence of a handful of theories already on the origin of the 1/\textit{f} range, a potentially important effect is not fully accounted for as of yet. The works cited above, except \citet{2016Ap&SS.361..364G}, are neglecting the effect of transverse structuring or inhomogeneity in the solar corona and solar wind, as an additional property in shaping the large-scale dynamics. In the recent years it has become increasingly clear that the solar corona and solar wind are far from being homogeneous. Comet Lovejoy, acting as a natural probe, plunged deep into the solar corona with a perihelion of a mere 140 Mm from the solar surface, revealing density contrasts of at least a factor of six between neighboring flux tubes over scales of a few Mm \citep{2014ApJ...788..152R}. Processed STEREO-A/COR2 images showed that the density in the corona out to 14 $R_\odot$ varies by an order of magnitude on spatial scales of 50 Mm \citep{2018ApJ...862...18D}. The measured large-scale density amplitude variations showed a 1/\textit{f} spectrum. In-situ measurements of the density power spectrum usually reveal a -5/3 slope at 1 AU \citep{1990JGR....9511945M,2005PhRvL..94t4502H}, thought to originate in the passive advection of the plasma by the underlying turbulence \citep[e.g.,][]{0004-637X-562-1-279}. However, there are also indications of steeper slopes, around -1.8 \citep{2015ApJ...803..107S}. Closer to the Sun, in-situ measurements by the Parker Solar Probe yield shallower density spectrum slopes of around -1.4 \citep{2020ApJS..246...44M}. Based on radio scintillation analysis, it appears that the density power spectrum undergoes a shift from the Kolmogorov -5/3 spectrum above 20 $R_\odot$ to a spectrum of -1 below it  \citep{1979JGR....84.7288W}. Based on the fine structure of observed type II radio bursts, the spectrum of density perturbations was shown to follow a power law of -5/3 or steeper, at spatial scales of up to 50 Mm, at a distance of 2 $R_\odot$ \citep{2021ApJ...921....3C}. Comparing the found density power spectral slopes cited above, there appears to be no agreement on a specific value closer to the Sun, the measured values ranging from -2 to -1. Subsequent measurements of the density spectrum closer to the Sun by the Parker Solar Probe might shed a light on the origin of the disagreements on the measured power spectral slopes. \par 
		Transverse waves in the solar corona were shown to be omnipresent \citep{2007Sci...317.1192T, 2015NatCo...6E7813M}. The existence of structuring across the magnetic field has a strong impact on the properties of these transverse waves \citep{2008ApJ...676L..73V}. In a homogeneous plasma, the transverse waves are Alfv\'en waves, propagating along individual magnetic field lines with the local Alfv\'en speed. On the other hand, in a transversely structured plasma, so-called `collective' modes appear, surface and body modes such as surface Alfv\'en modes\footnote{These transverse collective modes, driven mostly by magnetic tension, are variously referred to as Alfv\'enic, kink, or surface Alfv\'en waves throughout the literature. In this work, partly also following \citet{2009A&A...503..213G}, we prefer to use the term `surface Alfv\'en waves' instead of `kink waves', in order to highlight the non-axisymmetric nature of the surface waves existing on random inhomogeneities, as in the current setup.}, which are coherently propagating transverse displacements of a bundle of magnetic field lines with different Alfv\'en speeds. Therefore, while Alfv\'en waves undergo phase mixing in a structured plasma, that is, they propagate at different speeds on different individual field lines \citep{1983A&A...117..220H,2018ApJ...859L..17S}, the linear evolution of surface Alfv\'en waves is more complex \citep{1991ApJ...376..355P}, involving among other effects, resonant absorption \citep[e.g.,][]{1978ApJ...226..650I,2002ESASP.505..137G} and phase mixing around resonant layers \citep{2010ApJ...711..990P,2015ApJ...803...43S,2021A&A...648A..22D}. These linear processes contribute in shaping the perpendicular spectrum of waves as they cross the highly inhomogeneous solar corona. If the fluctuations in the 1/\textit{f} range indeed belong to the energy-containing scale of turbulence, linear processes might dominate in their evolution. Studying the effect of transverse inhomogeneity, through 3D MHD simulations, on the turbulent evolution of waves in the solar corona and wind, \citet{2021ApJ...907...55M} pointed out that the energy spectrum of the first few wavefronts scaled as -1, before evolving to a steeper turbulent slope of around -5/3. The authors attributed this initial 1/\textit{f} spectrum to the linear phase mixing of the transverse waves, being visible before the full development of a turbulent cascade. In that work, the transverse inhomogeneities were given by random Gaussian enhancements of density, without a specific power spectrum in mind, therefore it is not clear whether the power spectrum of the underlying density perturbations had a strong effect on the appearance of the 1/\textit{f} energy spectrum initially. \par
		In this paper, we look at the impact of transverse plasma structuring, modeled after the observed density spectra, on the propagation and evolution of linear transverse waves. In particular, we determine the energy spectra for different wave modes and background structuring that results from their propagation through such a plasma. We seek to determine how properties of the background influence the resulting spectra of the fluctuations. \par 
		This paper is structured as follows. In Section~\ref{sec:model}, the mathematical and numerical models are described, in Section~\ref{sec:result}, the results are presented, and in Section~\ref{sec:concl} we discuss the potential implications and conclusions of this study.
		
		\section{Model description} \label{sec:model}
		
		We use two different models to study the impact of transverse plasma structuring on the appearance of propagating transverse waves. These models will be described separately in two subsections in the following. A shared and key component for these two models is the transverse background profile itself, given by density variations. Different transverse density profiles are tested, based on the range of observed density power spectral slopes between -2 and -1. We generate the transverse profile in density using a two-dimensional (2D) colored noise generator. For this, we first generate a 2D white noise (i.e. flat power spectrum). Then, depending on the chosen model for the density power law, the Fourier transform of the white noise 2D map is multiplied with the corresponding normalized power law. Using an inverse Fourier transform, the 2D noise profile is obtained. In our analysis we tested Brownian (power spectrum of -2), Kolmogorov (-5/3), and pink (-1) noises for the underlying density fluctuations.
		
		\subsection{Alfv\'en wave model}
		
		The first model is a simple Alfv\'en wave propagator, through which we aim to demonstrate the shortcomings of models in which the background plasma inhomogeneity is not self-consistently accounted for. In other words, we treat the transverse waves as pure Alfv\'en waves, without taking into account the full MHD coupling of the perturbations to the inhomogeneous background \citep[e.g.,][]{2019FrASS...6...20G}. In this approximation, each magnetic field line is considered to be independent, and supporting a propagating Alfv\'en wave. Therefore, the full solution is described by the collection of individual Alfv\'en wave `rays'. This is similar to the WKB approach employed in \citet{2013A&A...549A..54M}. \par
\begin{figure*}[t]
			\centering     
			\begin{tabular}{@{}ccc@{}}
				\includegraphics[width=0.33\textwidth]{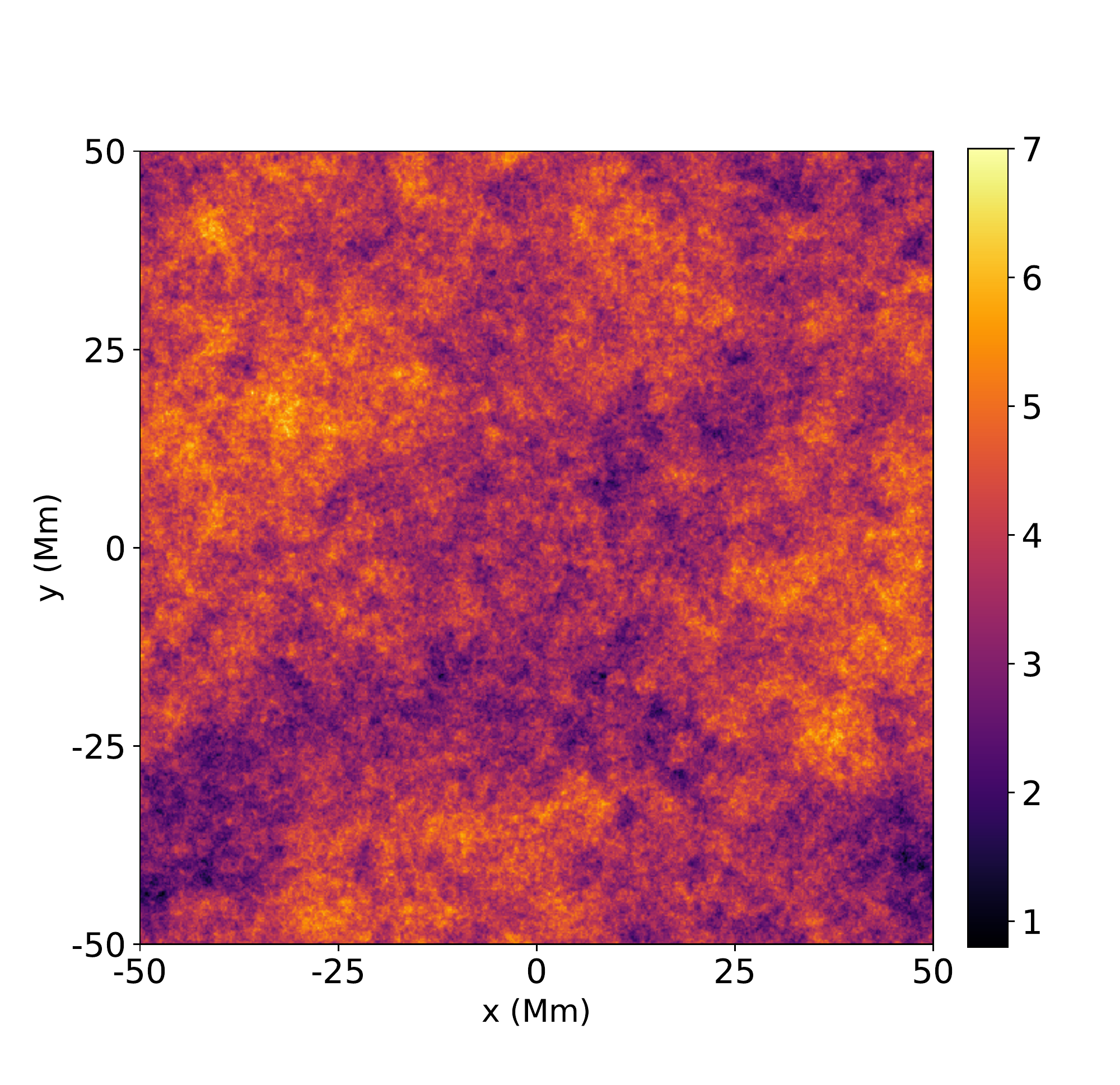}  
				\includegraphics[width=0.33\textwidth]{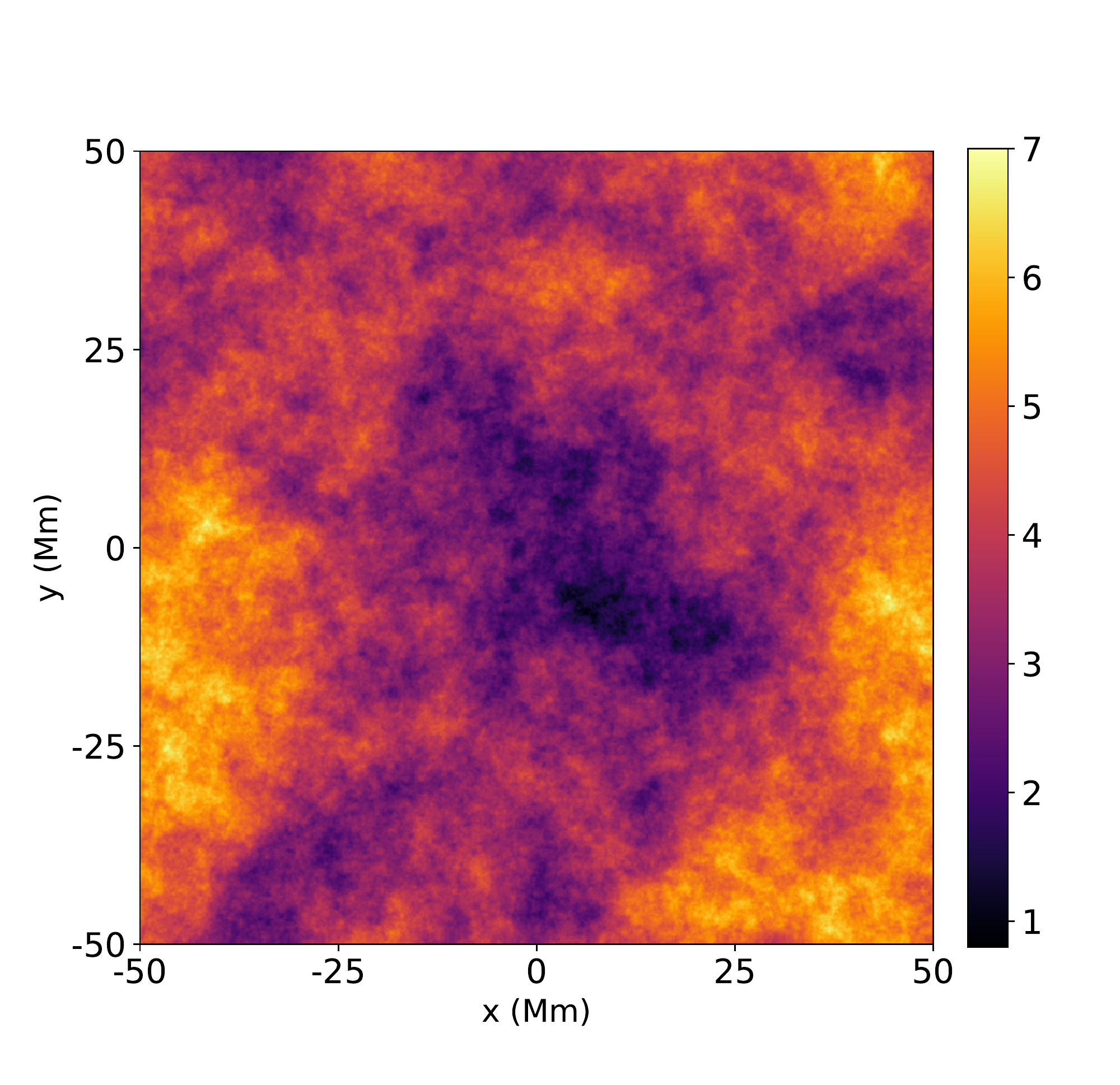}
				\includegraphics[width=0.33\textwidth]{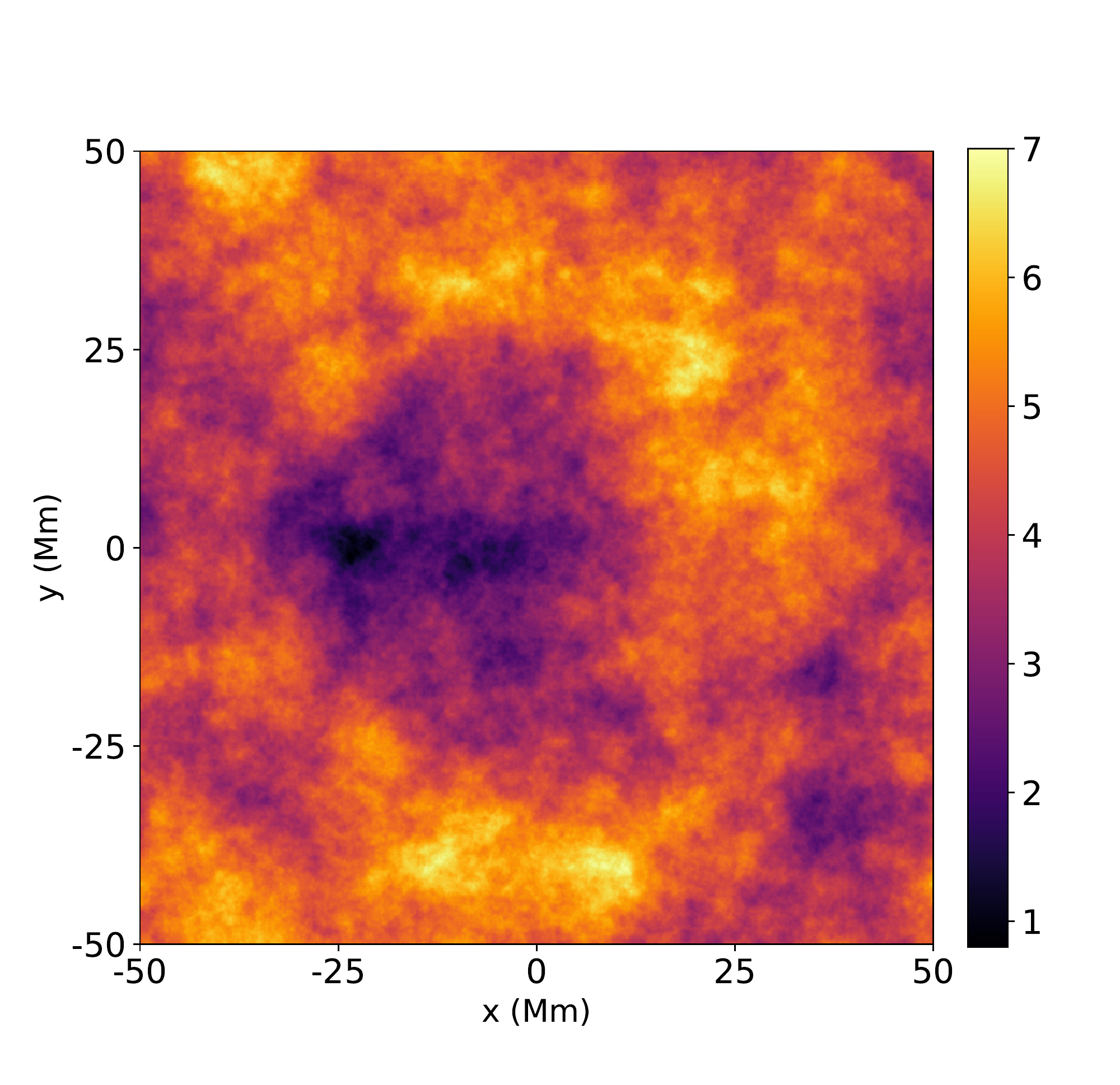}
			\end{tabular}    
			\caption{Two-dimensional noise maps used to generate the transverse density inhomogeneity for the numerical analysis. For the surface Alfv\'en wave setup, the maximum density is reduced to 2 in each plot. \textit{Left}: Pink noise ($f(k_\perp) \sim k_\perp^{-1}$). \textit{Centre}: Kolmogorov noise ($f(k_\perp) \sim  k_\perp^{-5/3}$). \textit{Right:} Brownian noise ($f(k_\perp) \sim  k_\perp^{-2}$).}
			\label{fig1}
		\end{figure*}

\begin{figure*}[h]
			\centering     
			\includegraphics[width=0.5\textwidth]{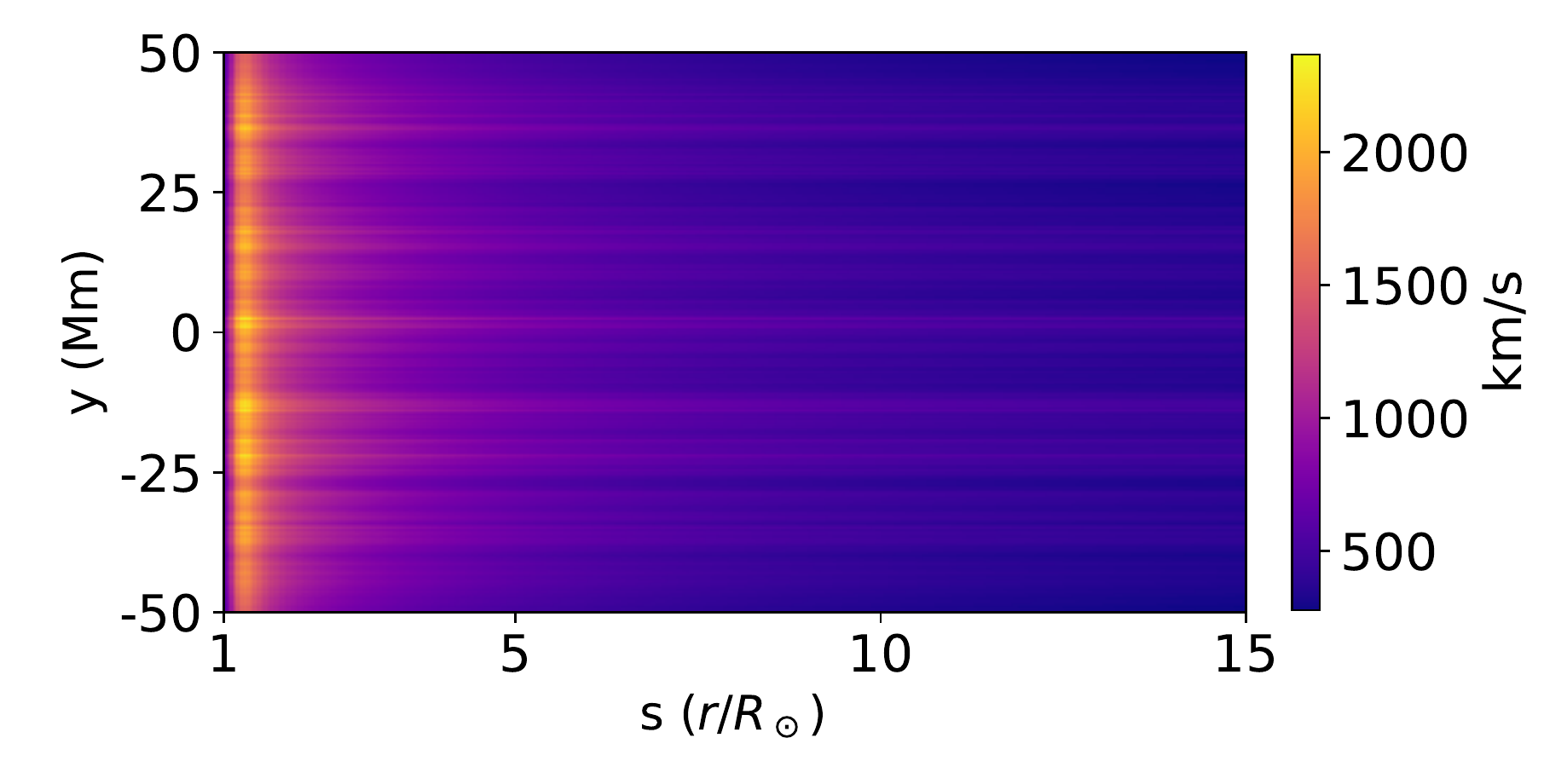}
			\caption{The sum of the Alfv\'en speed and background flow in a slice at $x=0$, for the Kolmogorov 2D noise map, for the Alfv\'en wave setup. The spatial extent along the $y$-direction is multiplied by $s$.}
			\label{fig2}
		\end{figure*}

 
		The background plasma is a 3D domain based on a 2D noise map described above (see Fig.~\ref{fig1}), which is multiplied by the base density value of $3.25 \cdot 10^8\  \mathrm{cm^{-3}}$. The density map obtained acts as the bottom radial boundary condition, and is considered to be the base of the corona. The extent of the 3D domain in the radial direction is 30 $R_\odot$, while in the horizontal directions it is 100 Mm at the base of the corona. The 3D domain is considered to be an `expanding box', also called a `narrow flux tube' approximation, in which $x$ and $y$ are Cartesian components in the plane perpendicular to the radial magnetic field \citep[e.g.,][]{2013ApJ...776..124P}. This approximation holds if $x^2 + y^2 \ll r^2$, and our analysis can be considered as leading-order in an expansion in powers of $\theta_{max} \approx \mathrm{arccos}(x/r)$, in a spherical coordinate system. The discretization of the numerical domain is $512^3$ cells. Starting from the bottom radial boundary, each radial line is populated with an analytical solar wind solution \citep{2009ApJ...707.1659C}, which gives the full 3D inhomogeneous background plasma:
		\begin{equation}
			\begin{array}{ll}
				\rho(x,y,s) &= A_{2D}(x,y) \left( \frac{3.23 \times 10^8}{s^{15.6}} + \frac{2.51 \times 10^6}{s^{3.76}} + \frac{1.85 \times 10^5}{s^2}\right)\ \mathrm{cm^{-3}}, \\
				B_0(s) &= \frac{1.5}{s^2}\ \mathrm{G},
				\\
				U(x,y,s) &= 9.25 \times 10^7 \frac{B_0(s)}{\rho(x,y,s)}\ \mathrm{km/s},
			\end{array}
			\label{IC}
		\end{equation}
		where $s$ is r/$R_\odot$, with $R_\odot$ being the solar radius (see Fig.~\ref{fig2}), and $A_{2D}(x,y)$ is the 2D noise map.
		The obtained background model is populated with solutions of propagating pure Alfv\'en waves along each magnetic field line. The one-dimensional equation describing the propagation of Alfv\'en waves is obtained from the incompressible MHD equations:
		\begin{equation}
			\begin{array}{ll}
				\frac{D \rho}{D t} &= 0 \\
				\frac{D \mathbf{v}}{D t} &= -\frac{1}{\rho} P_T + \frac{1}{\mu \rho} (\mathbf{B} \cdot \nabla) \mathbf{B}, \\
				\frac{D \mathbf{B}}{D t} &= - (\mathbf{B} \cdot \nabla) \mathbf{v},
			\end{array}
			\label{MHDeq}
		\end{equation}
		where we implied the solenoidality of both $\mathbf{v}$ and $\mathbf{B}$, and $D/Dt = \partial/\partial t + \mathbf{v} \cdot \nabla$ is the convective derivative. Incompressibility is a valid assumption as we are interested in the dynamics of linear Alfv\'en waves. For the study of Alfv\'en waves it is convenient to switch to \citet{1950PhRv...79..183E} variables, defined as $\mathbf{z}^\pm = \mathbf{v} \pm \mathbf{B}/\sqrt{\mu \rho}$. Writing the Els\"{a}sser variables as a sum of background and perturbed quantities, $\mathbf{z^\pm} = \mathbf{z_0^\pm} + \mathbf{{z_\perp^\pm}'}$, the equations for first order perturbations (with dropped indices and subscript) after some algebra read \citep{1980JGR....85.1311H,2005ApJS..156..265C}
		\begin{equation}
			\frac{\partial z^\pm}{\partial t} + (U \mp V_A) \frac{\partial z^\pm}{\partial r} = (U \pm V_A) \left(\frac{z_\pm}{4 H_D} + \frac{z_\mp}{2 H_A}\right),
			\label{Elseq}
		\end{equation}
		where $V_A = B_0/\sqrt{\mu \rho}$ is the Alfv\'en speed, $H_D = \rho/(\partial \rho/\partial r)$ and $H_A = V_A/(\partial V_A/\partial r)$ are the scale heights for density and Alfv\'en speed, respectively. We note that $H_A$ couples the outward and inward propagating Alfv\'en waves, causing their continuous reflection. By convention, $z^+$ represents outward propagating Alfv\'en waves. Next, we apply a WKB expansion on Eqs.~\ref{Elseq} with the smallness parameters defined by the ratio of scale heights to wavelength, $\epsilon = 1/(k H)$. The leading order expansion yields the dispersion relation of Alfv\'en waves \citep{1973JGR....78.3643H}:
		\begin{equation}
			\omega = k(U \mp V_{A}),
		\end{equation}
		where $\omega$ and k are the wave frequency and wavenumber, respectively.
		The next order of the WKB expansion yields the energy conservation of the propagating Alfv\'en waves \citep{1973JGR....78.3643H}:
		\begin{equation}
			z^+ \sim \rho^{1/4} \frac{U}{(U+V_A)}.
		\end{equation}
		Higher order terms in the WKB expansion allow for the reflection of outward-propagating Alfv\'en waves \citep{1990JGR....9514873H}, which we neglect here. Reflection or linear coupling of Alfv\'en waves is important in nonlinear studies, e.g. Alfv\'enic turbulence generation \citep[e.g.,][]{1999ApJ...523L..93M,2007ApJS..171..520C,2009ApJ...700L..39V}. However, here we only consider the linear evolution of Alfv\'en waves. The WKB approximation was shown to work well in the solar wind even for the low frequency waves studied here \citep[e.g.,][]{1980JGR....85.1311H,2013LRSP...10....2B}. For the present model, with $f$ = 1 mHz, $\epsilon < 1$ at $r \gtrsim 2\ R_\odot$, and $\epsilon < 0.1$ at $r \gtrsim 12\ R_\odot$. Numerically solving Eq.~\ref{Elseq} also demonstrates the validity of the WKB approximation. Therefore, we do not expect reflected waves to have a significant contribution in the resulting wave spectra. 
		The Alfv\'en wave solution is then expressed as
		\begin{equation}
			\mathbf{z^+}(x,y,s) = A\ \rho(x,y,s)^{1/4}\frac{U(x,y,s)}{(U+V_A)(x,y,s)} \mathcal{F}\left( s\frac{\omega}{(U+V_A)(x,y,s)}\right),
			\label{Alfveq}
		\end{equation}
		where $\mathcal{F}$ is the envelope function of the continuously driven Alfv\'en waves, here taken to be the cosine function. $A$ is an amplitude normalization term, which ensures that $\mathbf{z^+}(x,y,s) = 10\ \mathrm{km/s}$ at $s = 1$. The angular frequency $\omega = 2 \pi f$ is chosen such that the frequency is within the observed 1/\textit{f} range in the solar corona \citep{2021PhRvL.127y5101K}, taking two values, $f$ = 0.5 and 2 mHz. We note that as a consequence of the Alfv\'en wave solution as expressed above, at the bottom radial boundary the waves have an infinite transverse correlation length, or in other words, they are in phase across the transverse directions. Although the transverse correlation length of fluctuations is not known at the base of the corona, it is not expected to be larger than the supergranulation scale of around 30 Mm \citep[e.g.,][]{2003ApJ...597.1200R}. Therefore, the complete correlation of the signal at the bottom radial boundary is a `worst case scenario', in the sense that any transverse structuring of the waves is induced by phase mixing alone. Eq.~\ref{Alfveq} has no temporal dependence, instead representing the wave solution for a specific initial phase, in the whole domain. Altering this initial phase does not affect the resulting spectra.
		
		\subsection{Surface Alfv\'en wave model}
		\label{surface_desc}
		
		The second model presented in this work takes into account the full coupling of the fluctuations to the transversely inhomogeneous background plasma, by solving the full 3D compressible MHD equations numerically. For these simulations we use the \texttt{FLASH}\footnote{Website: https://flash.rochester.edu/site} code \citep{Dubey2013}. The second-order unsplit staggered mesh method is used, with Roe's solver and `mc' slope limiter. The method of constrained transport is used to ensure the solenoidality of the magnetic field. The 3D Cartesian numerical box is made up of $512^3$ cells, with spatial extents of $L_\perp =$ 100 Mm in the transverse directions and $15\ R_\odot$ in the radial direction. Additionally, we ran a simulation extending up to $30\ R_\odot$. The 2D noise map used to generate the transverse density profile is the same as in the previous model, albeit with the maximum magnitude reduced to 2.  
		However, in this numerical model we neglect radial variations and we consider no background flow. Therefore, the 3D domain is populated by repeating the obtained 2D density map in the third direction. Expansion is neglected. The background magnetic field of 12 G is straight and uniform, with plasma $\beta \approx 0.28$. At the bottom radial boundary, transverse velocity perturbations are induced using a time and space-varying driver used previously by \citet{2021ApJ...907...55M}. The wave driver adds a solenoidal velocity field at selected perpendicular wavenumbers following an Ornstein-Uhlenbeck process with a finite autocorrelation time, adapted from \citet{2010A&A...512A..81F}. Perpendicular wavenumbers from 1 up to 4 $k_0$ are stirred, where $k_0 = 2 \pi/L_\perp$, with all modes having the same energy, i.e. a flat spectrum. The autocorrelation time is set to $\tau_d = 500\ s$, corresponding to one of the same frequencies as in the previous setup. The amplitude of the perturbations is kept at a low level, in order to study only linear dynamics, with $v_{RMS} \approx 0.002\ V_{A_{min}}$. Besides the conditions for transverse velocities, parallel velocity is set to obey antisymmetry at the bottom radial boundary, ensuring that no strong secondary flows along the field lines are generated at the driven boundary. Other variables obey a zero-divergence or `open' condition at this boundary. The numerical box is periodic laterally, while at the top radial boundary all variables are set to zero-divergence. The reflection on the upper boundary is minimal, and does not affect the spectral properties of the waves. The total simulation time is one Alfv\'en crossing time relative to the smallest Alfv\'en speed in the domain. 
		
		\section{Results} \label{sec:result}
		
		\subsection{Alfv\'en wave propagation}
		
\begin{figure*}[h]
			\centering     
			\begin{tabular}{@{}cc@{}}
				\includegraphics[width=0.5\textwidth]{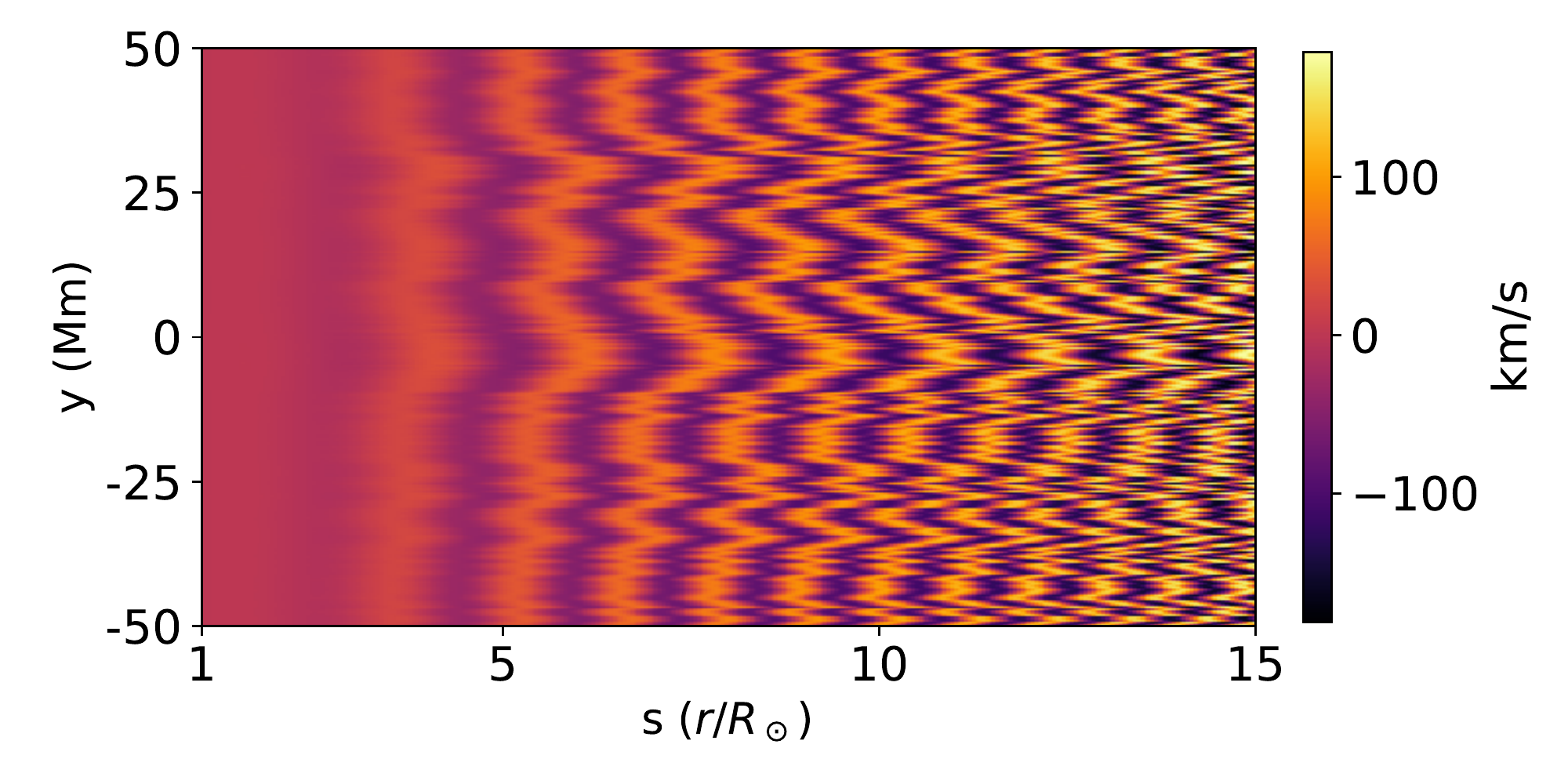}  
				\includegraphics[width=0.5\textwidth]{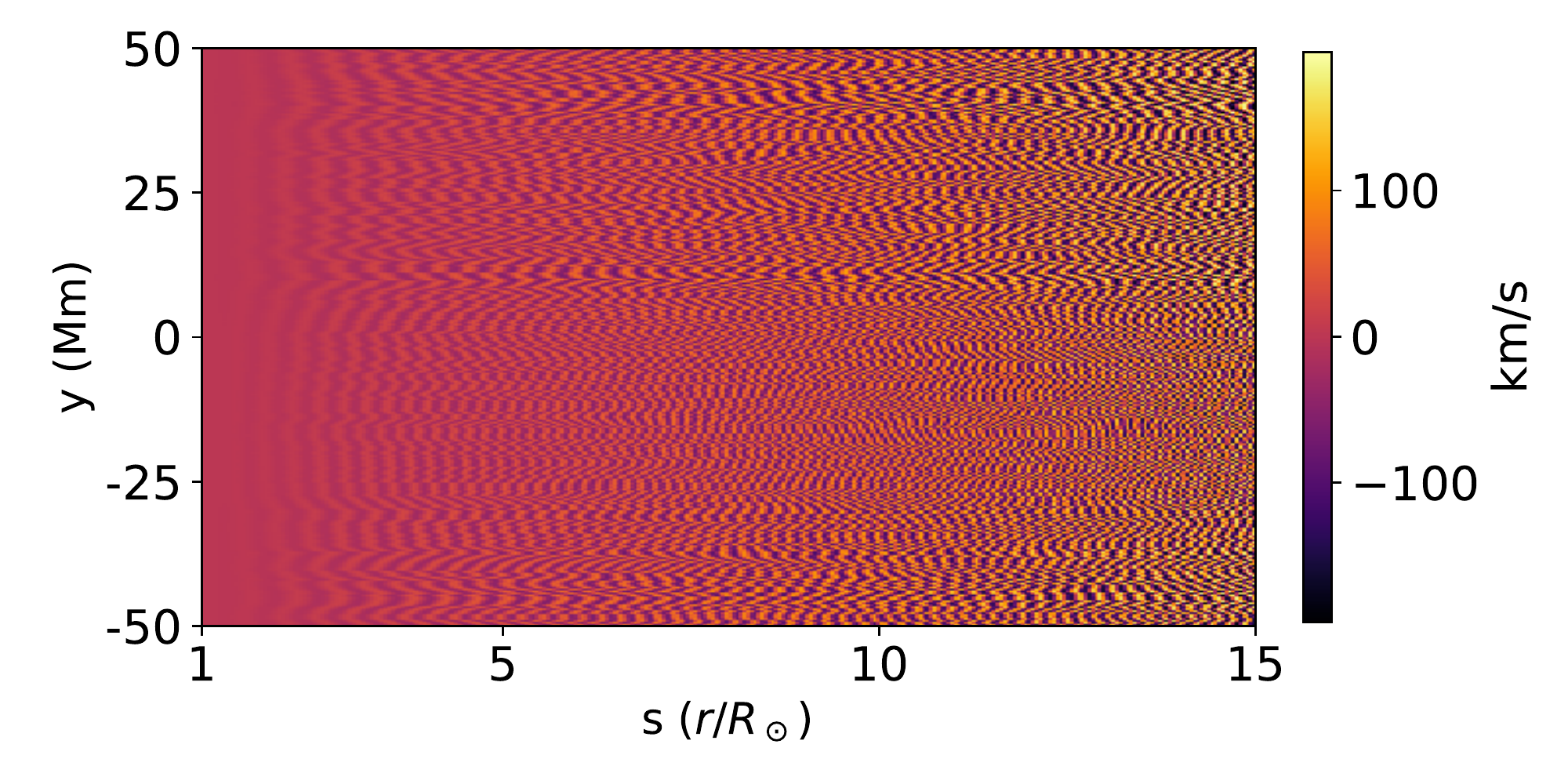}
			\end{tabular}    
			\caption{The Alfv\'en wave solutions of Eq.~\ref{Alfveq} in a slice at $x = 0$ for the Kolmogorov 2D noise map. \textit{Left:} f = 0.5 mHz or T = 2000 s. \textit{Right:} f = 2 mHz or T = 500 s. The spatial extent along the $y$-direction is multiplied by $s$.}
			\label{fig3}
		\end{figure*}

\begin{figure*}[h]
			\centering     
			\begin{tabular}{@{}cc@{}}
				\includegraphics[width=0.33\textwidth]{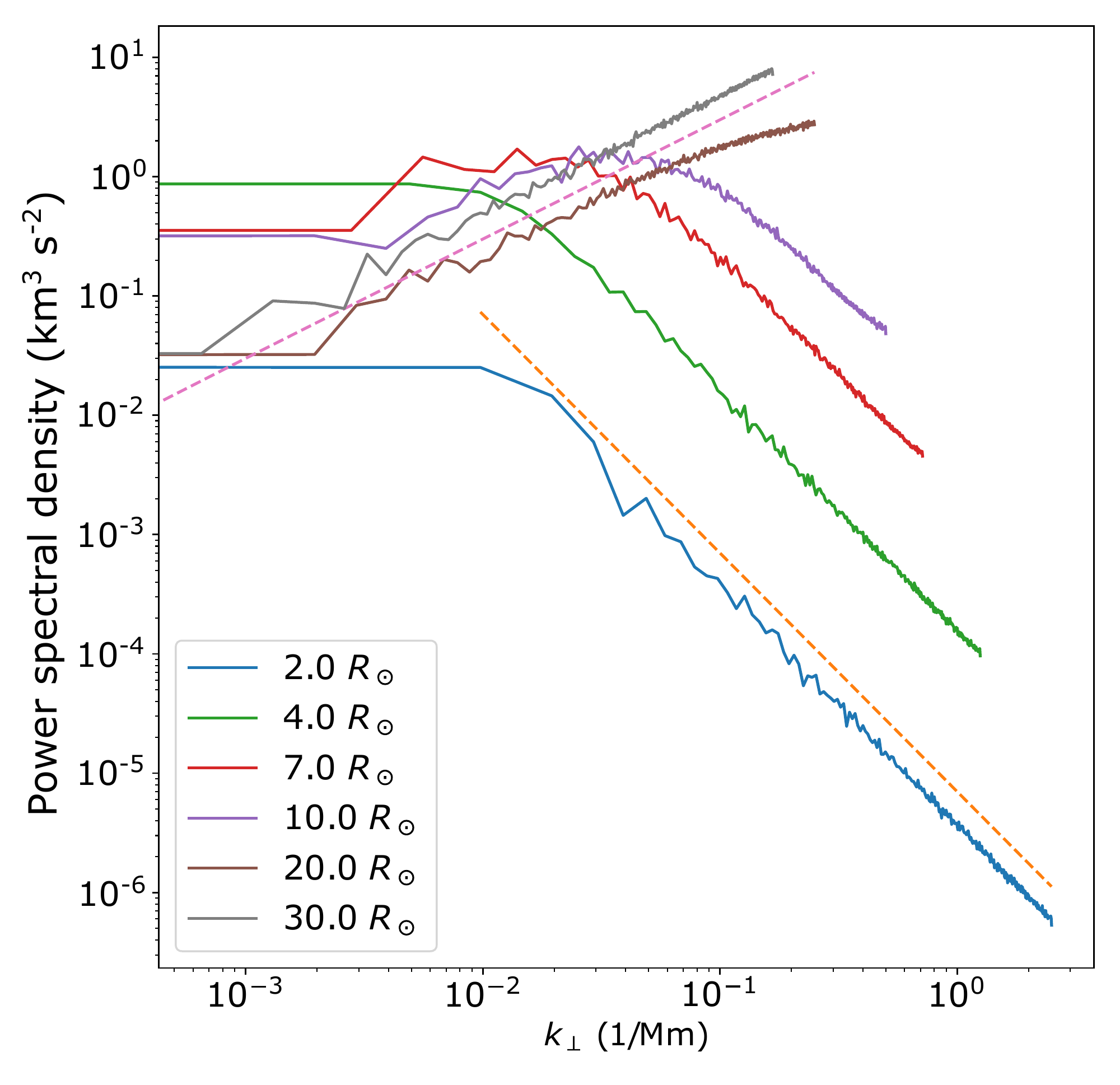}  
				\includegraphics[width=0.33\textwidth]{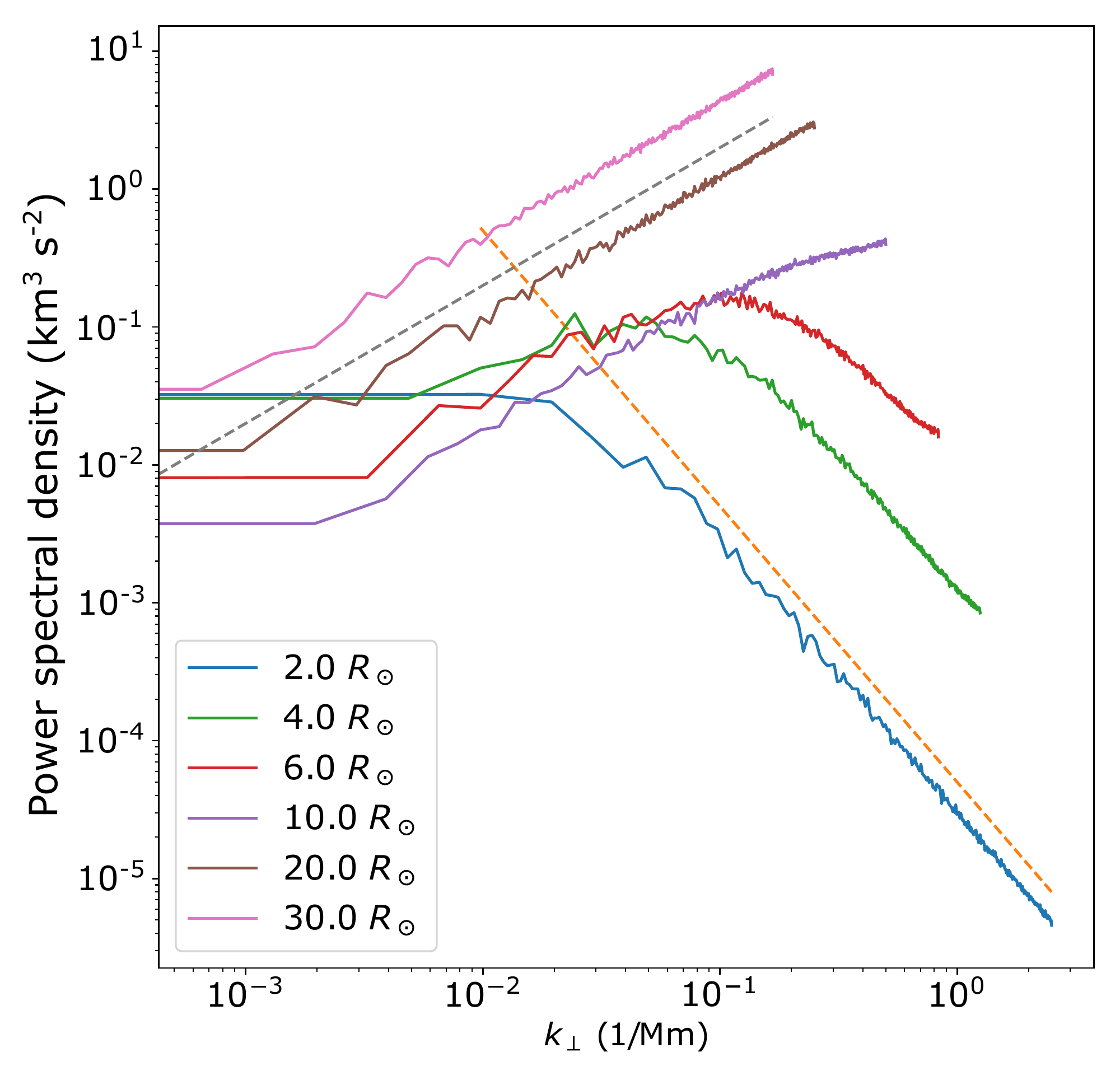}
			\end{tabular}    
			\caption{Alfv\'en wave perpendicular energy spectra at different heights, for the Brownian 2D noise map. \textit{Left:} f = 0.5 mHz. \textit{Right:} f = 2 mHz. The two dashed lines represent spectral slopes of -2 and 1.}
			\label{fig4}
		\end{figure*}

\begin{figure*}[h]
			\centering     
			\begin{tabular}{@{}cc@{}}
				\includegraphics[width=0.33\textwidth]{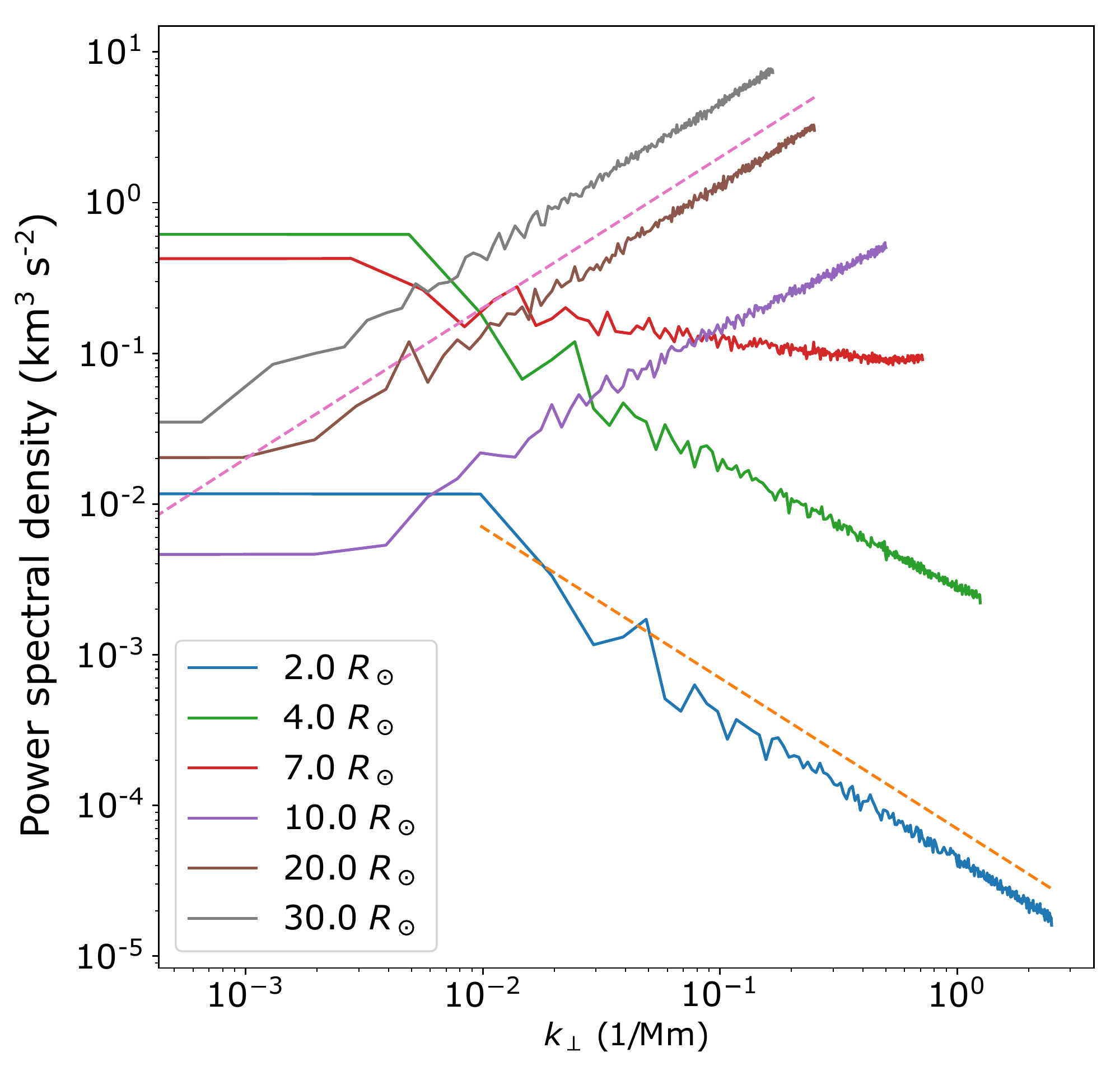}  
				\includegraphics[width=0.33\textwidth]{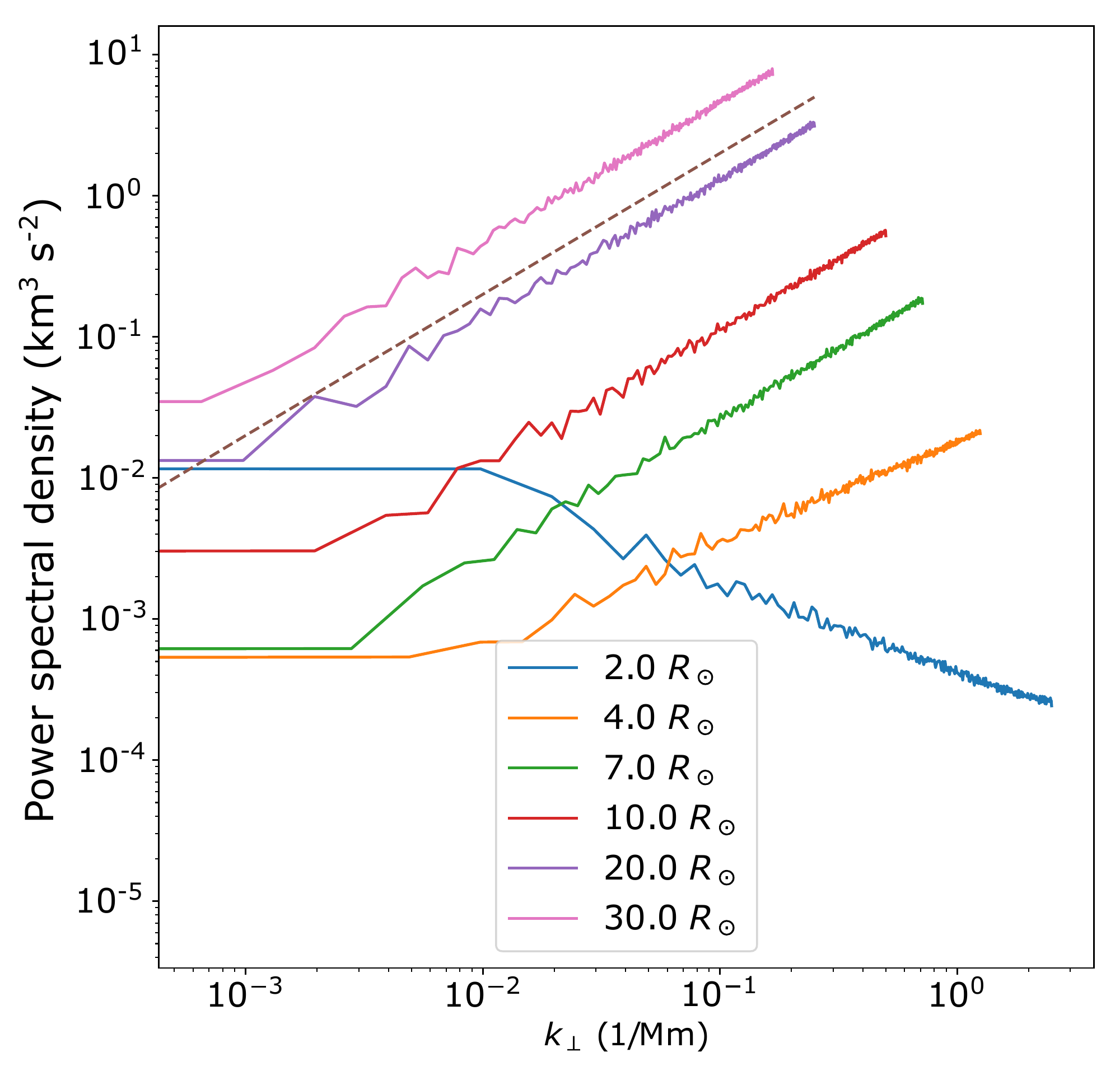}
			\end{tabular}    
			\caption{Alfv\'en wave perpendicular energy spectra at different heights, for the pink 2D noise map. \textit{Left:} f = 0.5 mHz. \textit{Right:} f = 2 mHz. The two dashed lines represent spectral slopes of -1 and 1.}
			\label{fig5}
		\end{figure*}

		The time-independent Alfv\'en wave solution in Eq.~\ref{Alfveq} is calculated for the background plasma equilibrium using the \texttt{numpy} \citep{harris2020array} library in \texttt{Python}. Vectorization through \texttt{numpy} offers a significant speed-up for the required element-wise matrix multiplications at this step. The obtained Alfv\'en wave solutions are illustrated in Fig.~\ref{fig3} in a cross-sectional slice of the full 3D map, for two different frequencies. In these plots, the main effects of the inhomogeneous Alfv\'en speed of the background can immediately be identified, both along and across the magnetic field direction. Along the magnetic field ($s$-axis), the amplitude is increasing with radial distance from the Sun, here described by the first order approximation of the WKB expansion. Across the magnetic field phase mixing is visible, which leads to increasingly more oblique wavefronts, depending on the local Alfv\'en speed gradient. A spectral analysis is performed at different 2D cross-sections along the $s$-direction. The perpendicular energy spectra is defined as
		\begin{equation}
			E(k_\perp) = \sum_{k_x} \sum_{k_y} |\mathbf{z}^+_{k_x,k_y}(s)|^2, \quad k_\perp = \sqrt{k_x^2+k_y^2},
			\label{PSD}
		\end{equation}
		where $\mathbf{z}^+_{k_x,k_y}(s)$ is the center-shifted (that is, the zero-frequency component is in the center) 2D discrete Fourier transform of the cross-section at $s$. This process is equivalent with integrating the center-shifted 2D Fourier map in the polar direction at different radii. The resulting spectra are shown in Fig.~\ref{fig4} and Fig.~\ref{fig5}. Initially, the perpendicular energy spectrum acquires the perpendicular power spectrum of the background density, irrespective of the density noise map used. The radial distance to which the wave energy spectrum can be approximated by the background power spectrum is a few wavelengths, but varies with the noise map. This observation implies that if transverse inhomogeneities are only present up to radial distances comparable to the wavelength of the waves, phase mixing alone would result in Alfv\'en waves acquiring the spectrum of the background. If the Alfv\'en waves continue to be phase-mixed, as in the present case, the perpendicular spectrum of the waves tends to evolve towards a positive slope of 1, again independent of the underlying noise map. A perpendicular power spectral slope of +1 can be interpreted as waves with equal energy over all wavelengths, or white noise along both perpendicular directions. In our analysis, a slope of +1 is recovered for both frequencies, and for all noise maps (with slopes from -1 to -2) by $20\ R_\odot$. Therefore, the phase mixing of independent Alfv\'en waves leads to a perpendicular power spectrum with a slope of +1, which does not depend on the underlying density inhomogeneity spectrum within the range of observed slopes. Energy spectra with positive slopes were as of yet not observed in the solar corona or wind. This finding is clearly the result of incomplete physics in the Alfv\'en wave model, which assumes the independence of Alfv\'en waves propagating on different magnetic field lines. More specifically, it shows that models which ignore the full coupling of MHD waves in the solar corona and solar wind to the inhomogeneous background fail to capture the correct transverse evolution of the waves in the linear regime. It can be argued that, by allowing nonlinear evolution in this model, the development of Alfv\'enic turbulence would render the power spectrum towards slopes of -1.5 or -5/3. If the interpretation of the spectral break observed in the solar corona and solar wind as the upper limit of the inertial range of turbulence is correct, on larger scales the power spectrum might be determined by linear processes. In this case, at the largest perpendicular scales, the power spectrum in an Alfv\'en wave model with included nonlinear dynamics would still be determined by phase mixing and display a slope of +1.  
		
		\subsection{Surface Alfv\'en wave propagation}
		
\begin{figure*}[h]
			\centering     
			\includegraphics[width=0.5\textwidth]{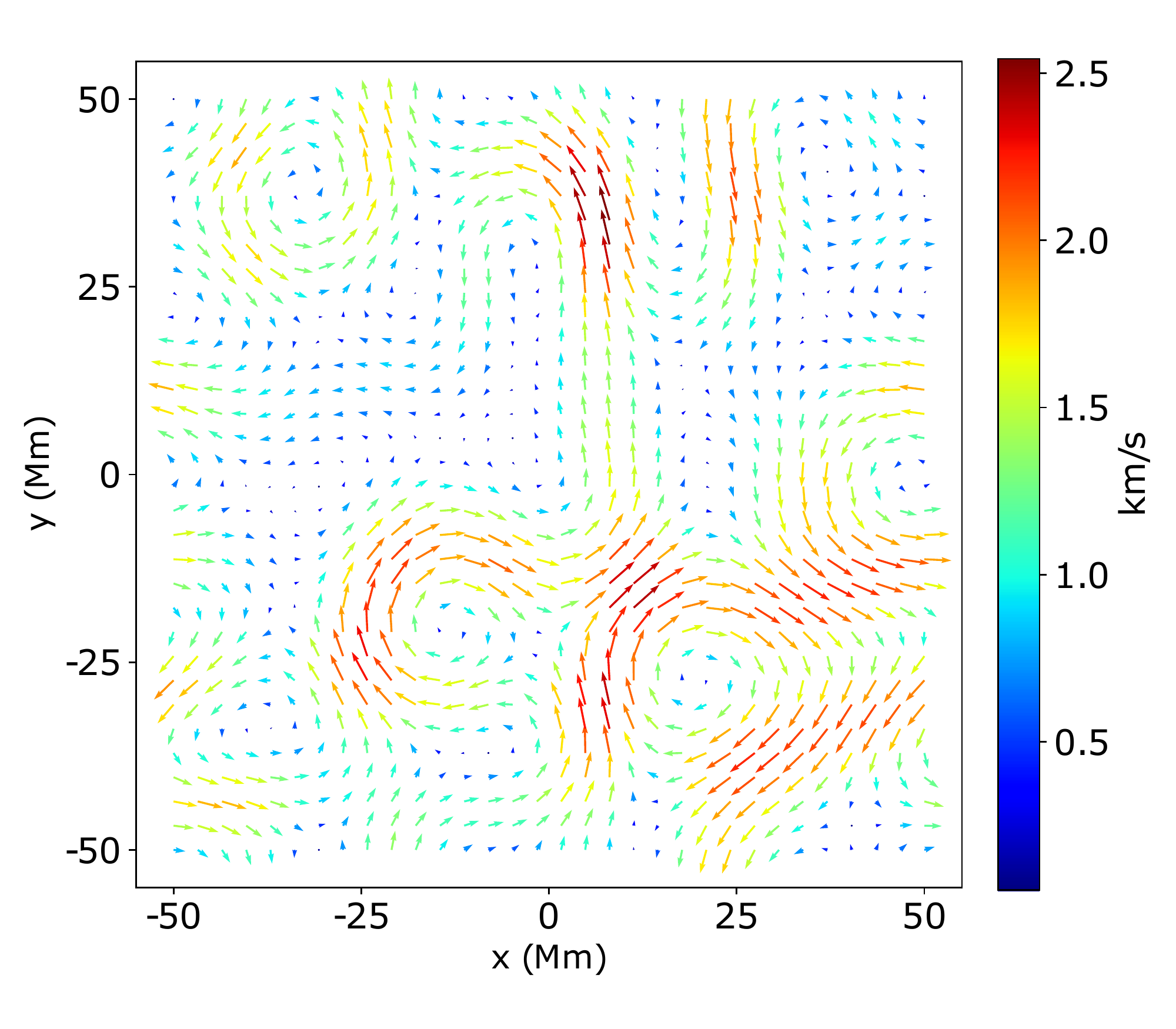}
			\caption{Vector plot of a single realization of the driven velocity perturbations at the bottom radial boundary, for the surface Alfv\'en wave setup. The color and size of the individual vectors represent velocity magnitude, in units of km/s.}
			\label{fig6}
		\end{figure*}

		As described in subsection~\ref{surface_desc}, the simulations run for one Alfv\'en transit time with respect to the slowest Alfv\'en speed in the system, $V_{A_{min}} \approx 0.5\ \mathrm{Mm/s}$. The wave driver injects transverse velocity perturbations at the bottom $s-$boundary (see Fig.~\ref{fig6}), and these displacements are picked up by the many flux-tube structures present due to the inhomogeneous density background profile. 
		
\begin{figure*}[h]
			\centering     
			\includegraphics[width=0.66\textwidth]{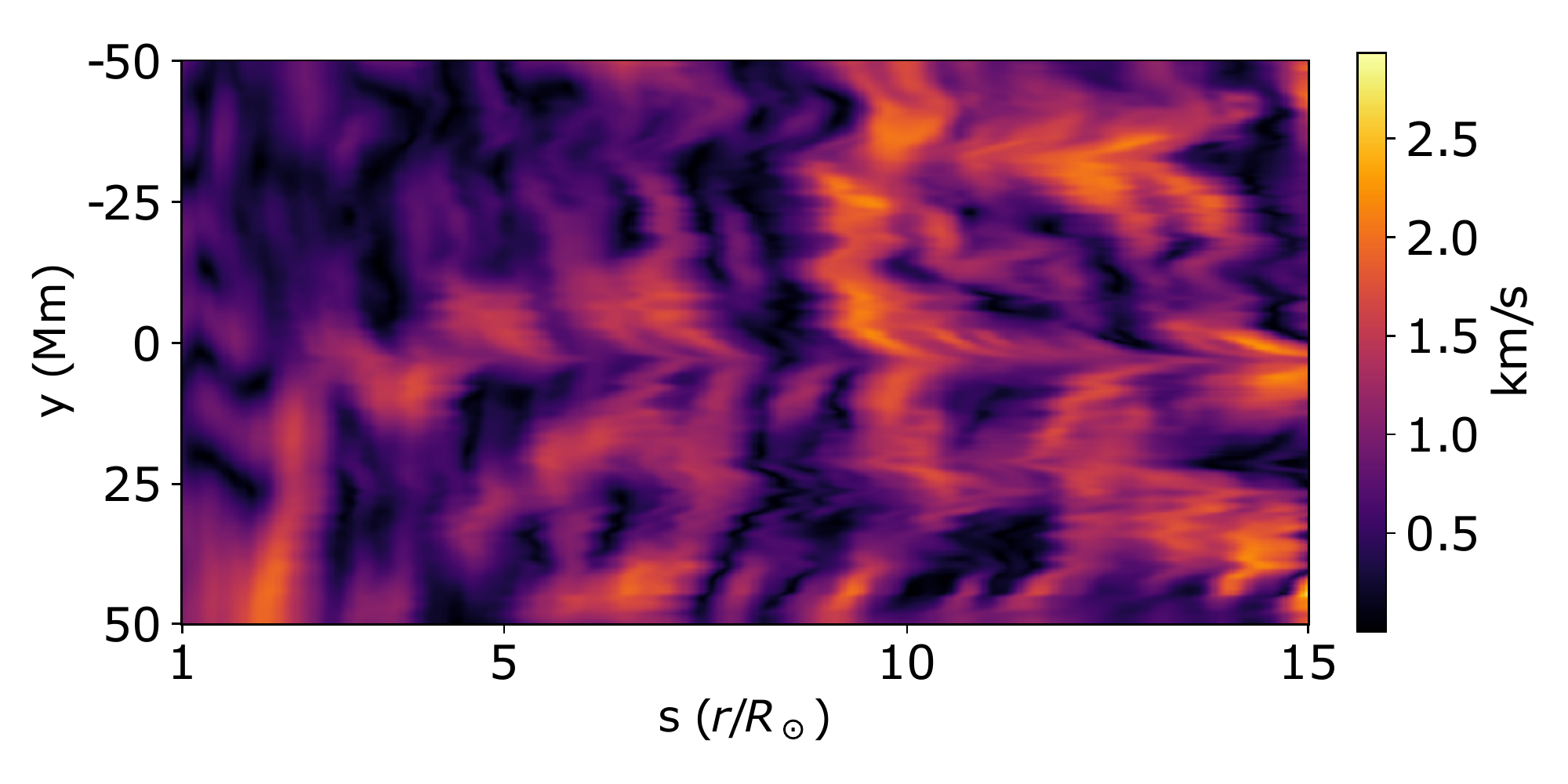}
			\caption{Plot of the perpendicular velocity perturbation magnitude in a slice along the background magnetic field at $x = 0$, at the end of the simulation time. The Kolmogorov noise density profile is used in this example.}
			\label{fig7}
		\end{figure*}

\begin{figure*}[h]
			\centering     
			\begin{tabular}{@{}ccc@{}}
				\includegraphics[width=0.33\textwidth]{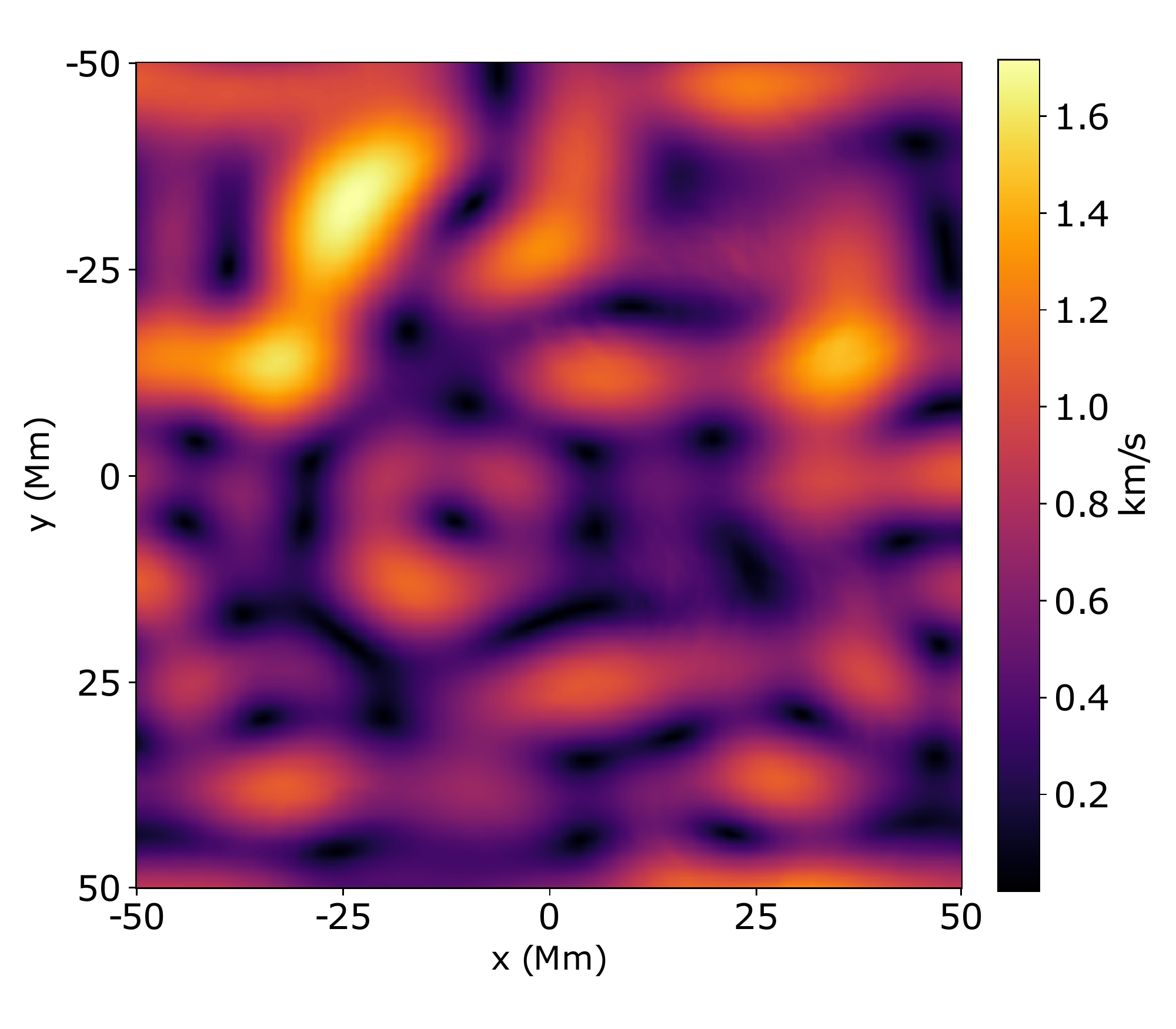}  
				\includegraphics[width=0.33\textwidth]{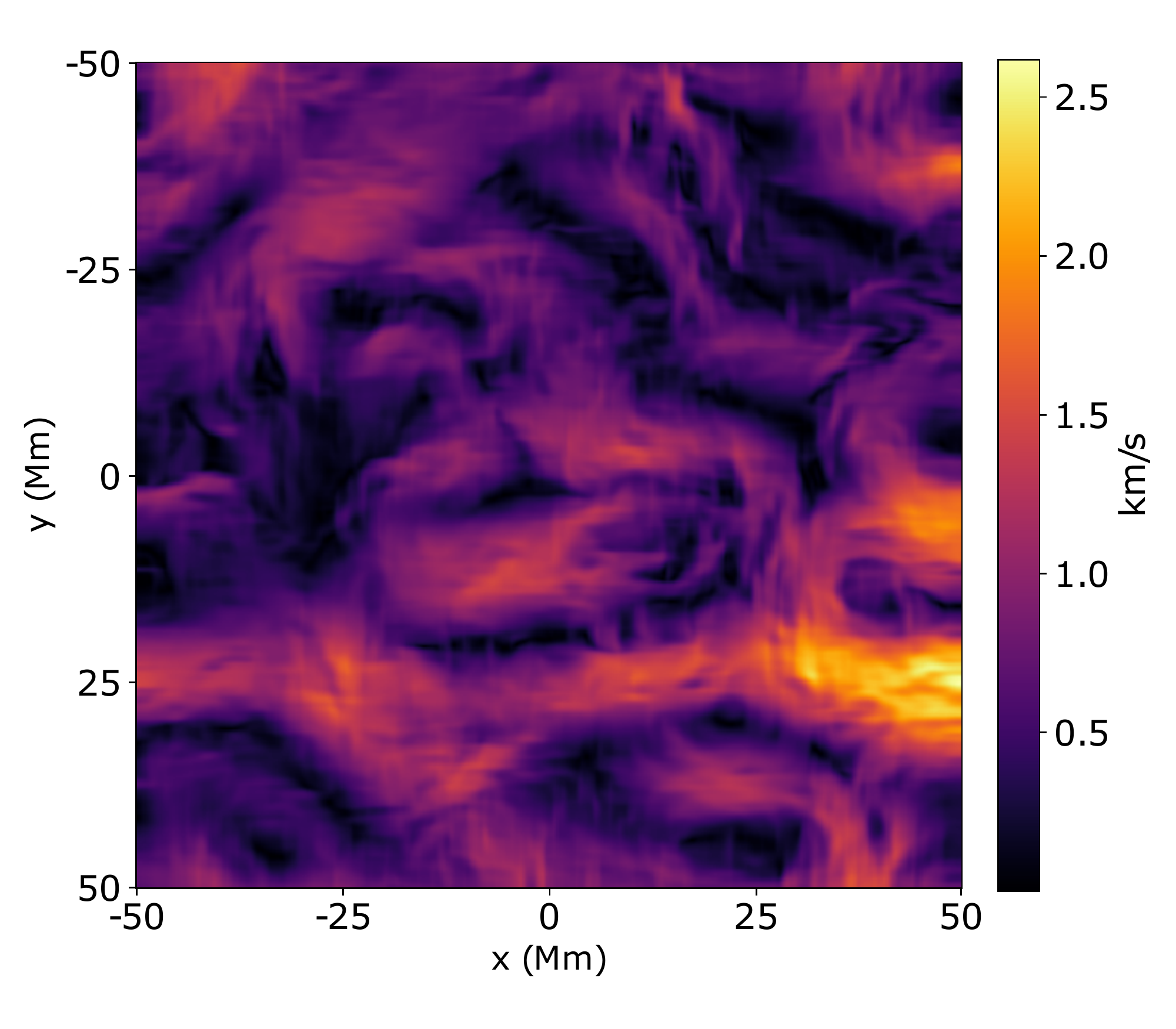}
				\includegraphics[width=0.33\textwidth]{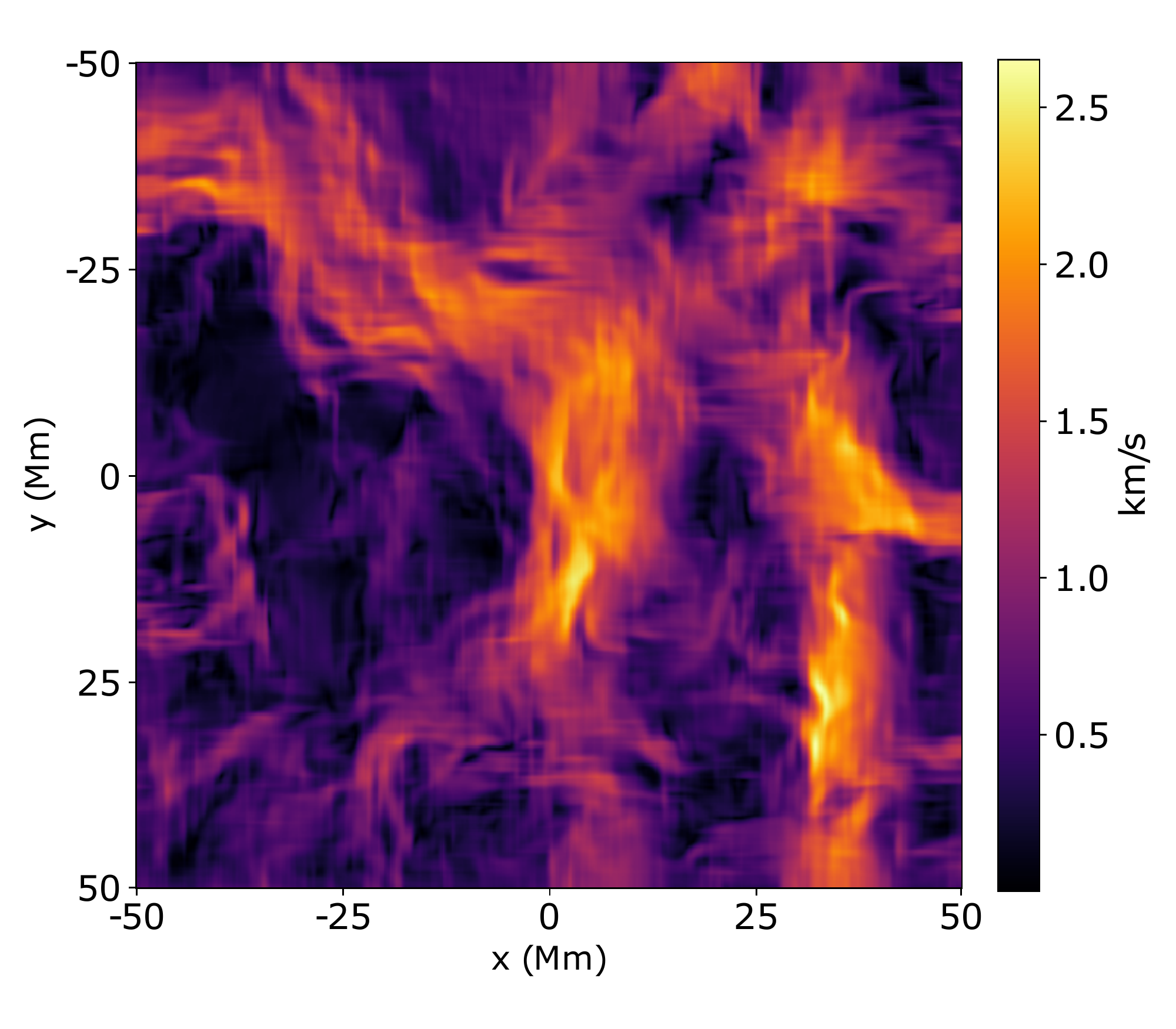}
			\end{tabular}    
			\caption{Plots of the perpendicular velocity perturbation magnitude in a slice across the background magnetic field, at three different radial distances, 1, 7.5, and 15 $r/R_\odot$, respectively (from left to right). The Kolmogorov noise density profile was used for this example. }
			\label{fig8}
		\end{figure*}

		Let us inspect the evolution of the waves in a slice along the radial direction, as depicted in Fig.~\ref{fig7}. 
		The development of small-scale structure across the magnetic field as a result of the linear processes described above can be observed. Gradients across the magnetic field appear increasingly steeper with increasing radial distance, noticeable also for the simple Alfv\'en wave rays undergoing phase mixing in Fig.~\ref{fig3}. However, as shown later, the energy spectrum displays an asymptotic behaviour rather than continuous increase at higher wavenumbers with increasing radial distance.
		In Fig.~\ref{fig8}, cross-sectional slices are shown at different heights. 
		
\begin{figure*}[h]
			\centering     
			\begin{tabular}{@{}ccc@{}}
				\includegraphics[width=0.33\textwidth]{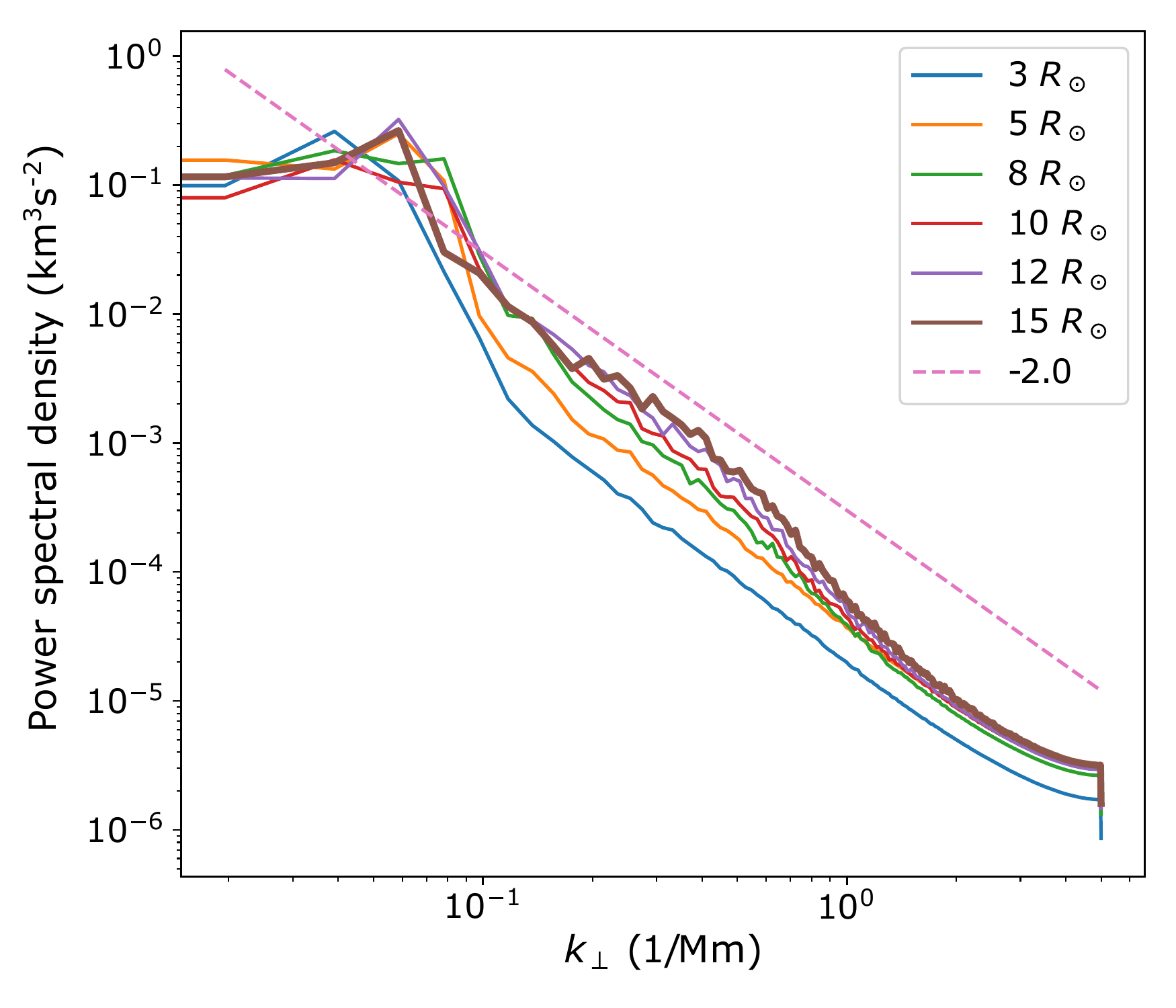}  
				\includegraphics[width=0.33\textwidth]{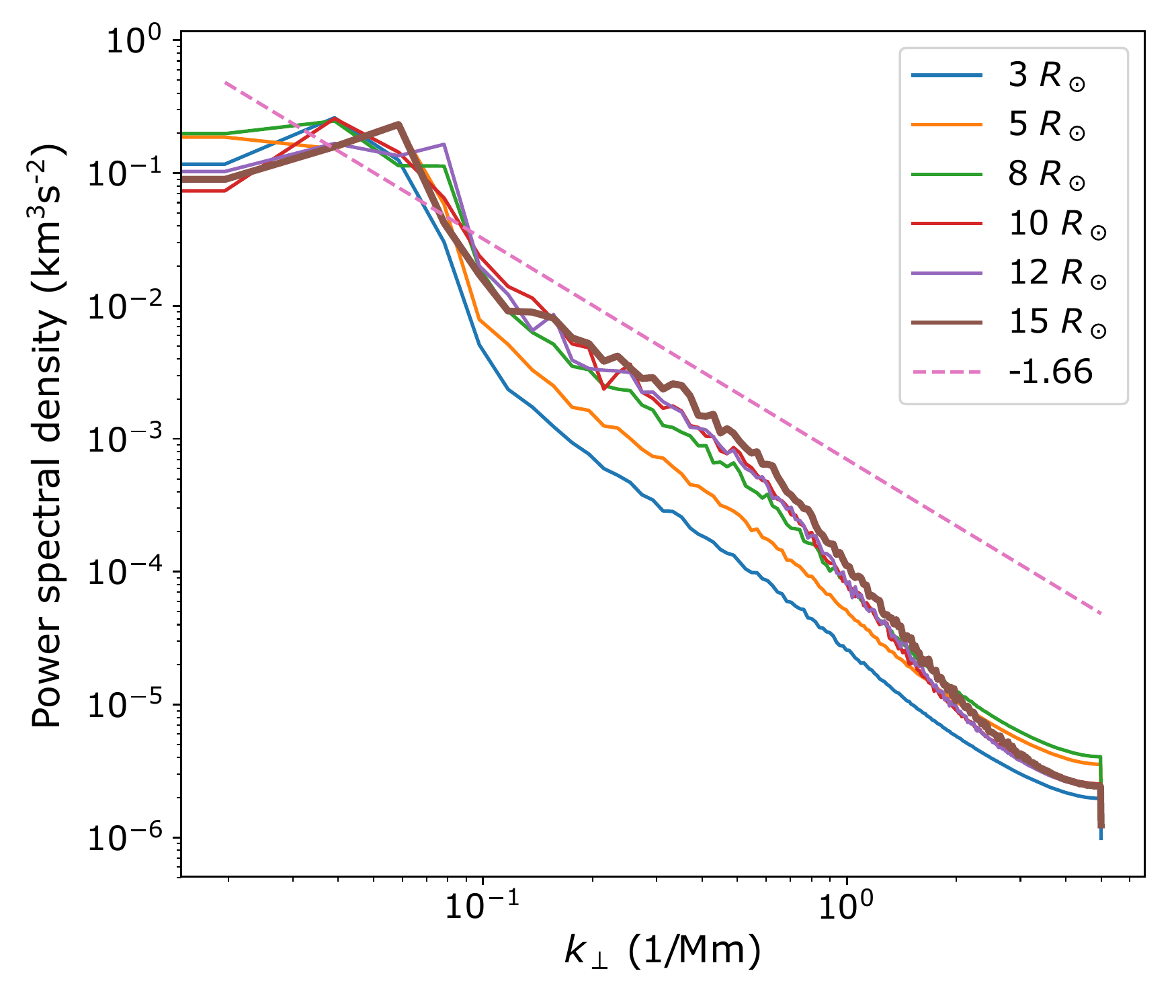}
				\includegraphics[width=0.33\textwidth]{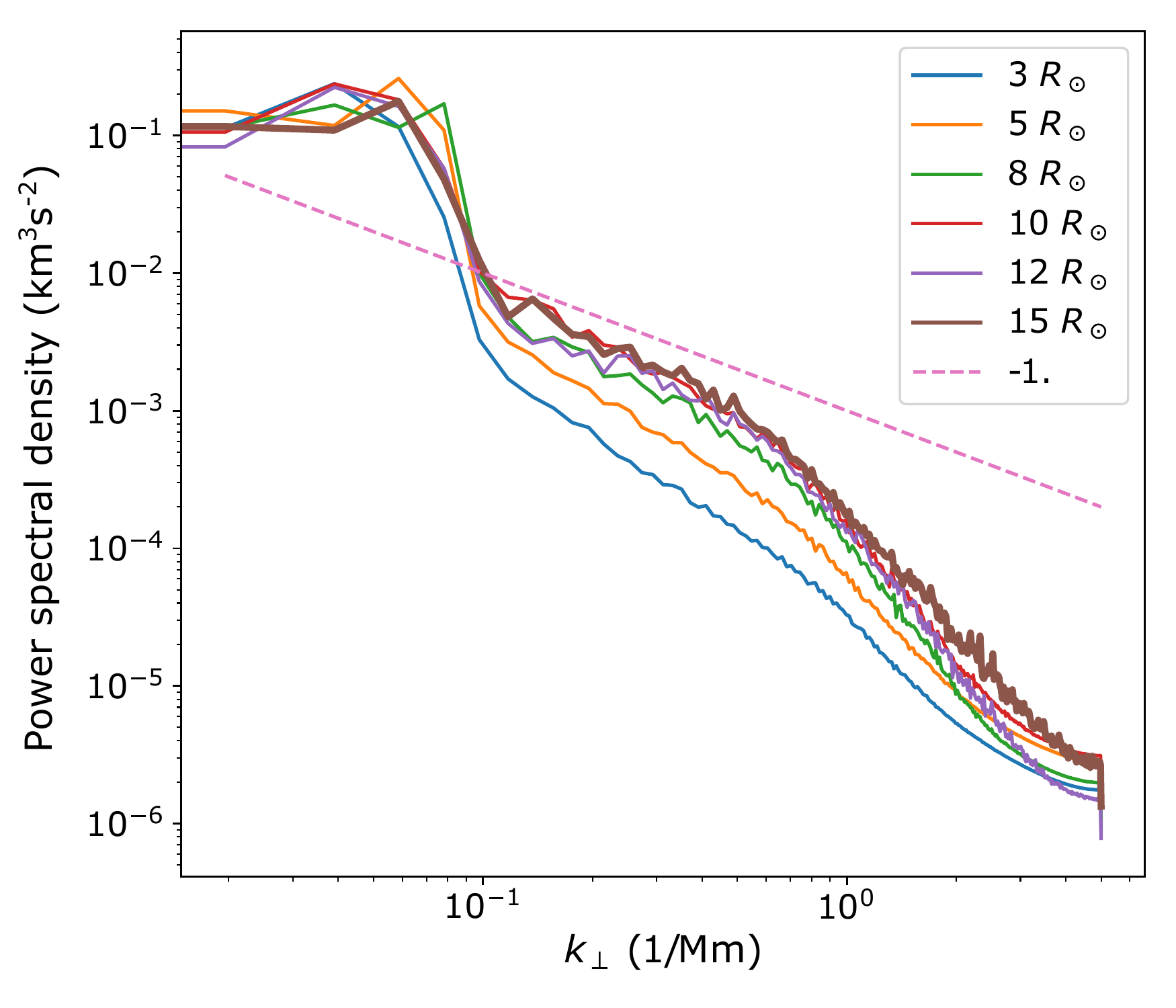}
			\end{tabular}    
			\caption{Perpendicular energy spectra at different heights, for the three noise maps used, brownian (-2), Kolmogorov (-1.66), and pink (-1), respectively (from left to right). The line corresponding to the spectra at 15 $R_\odot$ is thicker for easier identification. The dashed lines represent spectral slopes corresponding to the noise color.}
			\label{fig9}
		\end{figure*}

		The initial velocity perturbation induced by the driver is visible in the first image, showing the typical perpendicular wavenumbers perturbed. Smaller spatial scales appear in the next images, at larger radial distances. Yet, small wavenumbers, as induced by the driver, still appear dominant. In Fig.~\ref{fig9}, the evolution of the energy spectrum is shown for the three different density noise maps. 
		There are several observations that can be drawn from these plots. At large wavenumbers driven by the velocity driver, the evolution is minimal. A linear cascade is taking place towards smaller wavenumbers, which is depleting the energy at the driven wavenumbers. However, given the spatially and temporally evolving driver, the variation in the injected Poynting flux appears to be larger than the energy ending up at larger wavenumbers. The energy at larger wavenumbers than the driven ones appear solely as a result of the linear processes described above. We note that for each colored noise map, at these wavenumbers the energy spectral slope tends towards the slope of that particular density map. At even larger wavenumbers, the energy spectrum becomes steeper, arguably because of the numerical dissipation inherent to the solver. Convergence studies using the pink colored noise density map, out to 30 $R_\odot$, show that the energy spectrum does not become flatter with longer evolution, but that it does converge to the background slope value. Although we did not run a parametric study with regards to the autocorrelation time of the driver, the  linear processes acting here are proportional to the frequency of the waves. Therefore higher frequency waves converge towards the background noise spectrum slope faster. \par 
		Another important parameter is the density ratio employed, which directly controls the linear cascade rate. While no parametric study was carried out (for reasons described in Section~\ref{sec:concl}), a simulation with four times the maximal density ratio was completed. We find that the cascade from the driven scale is faster, and that it leads to a power law developing at the driven scales as well, in continuation of the power law at smaller scales. We also find that the energy spectrum still tends towards the spectrum of the background density map.  
		
		\section{Discussion and Conclusion} \label{sec:concl}
		
		The evolution of the observed 1/\textit{f} spectrum in the solar wind at low frequencies is still not entirely understood, despite the existence of a handful of theories aiming at explaining it. In this paper we have investigated an effect which was previously unaccounted for, that of background density inhomogeneities across the magnetic field. These inhomogeneities result in an Alfv\'en speed profile that is varying across different magnetic field lines, allowing for the existence of waveguides and collective waves propagating on them, such as surface Alfv\'en waves. To highlight the substantial difference between no coupling versus coupling of waves to such a varying background plasma, we study both the propagation of individual Alfv\'en waves and that of collective surface Alfv\'en waves, separately. We show that Alfv\'en waves undergoing phase mixing in a colored noise density background get completely out of phase over propagation distances that depend on their frequency and the underlying density map. This leads to a perpendicular energy spectral slope of +1, corresponding to random noise. Thus, the assumption that waves propagate independently through an inhomogeneous background on each magnetic field line as pure Alfv\'en waves, results in energy spectral slopes which are not observed in the solar wind. It could be argued that nonlinear processes that are not accounted for here would overpower the spectrum set up by the linear cascade and would lead to the well-known slopes observed in the inertial range of the solar wind turbulence. However, if the energy-containing low frequency part of the spectrum is not affected by nonlinear evolution, it would still display the +1 spectrum found here. \par
		In the second setup, we solve the full 3D MHD equations with velocity-driven boundaries, using the same density maps as in the Alfv\'en wave setup. The driven transverse displacements set up collective propagating oscillations of bundles of magnetic field lines, which are known as surface Alfv\'en waves (also referred to as kink or Alfv\'enic waves). The collective nature of these waves implies that different magnetic field lines or flux surfaces are not independent any longer, as in the case of individual Alfv\'en waves. This leads to a different behavior of these waves as they propagate through the transversely inhomogeneous plasma. The closely linked processes of phase mixing and resonant absorption \citep{2015ApJ...803...43S} linearly cascade wave energy from the driven scales to smaller scales, perpendicularly to the magnetic field. We investigate the evolution of the perpendicular energy spectrum with height for the different density maps. Energy builds up at spatial scales smaller than the driven wavenumbers, increasingly as a function of height. The energy spectrum slope at these scales converges towards the perpendicular power spectral slope of the density map used, for each of the three density map cases. The energy spectrum at the driven scales is decreasing with height in individual frames propagating with the waves, however, the decrease is small enough in the current setup to be hidden by the time variability of the energy injected by the wave driver. The main conclusion of these studies is that an inhomogeneous density with perpendicular power spectral slope of 1/\textit{f} can induce a linear perpendicular cascade of propagating surface Alfv\'en waves, the energy spectrum of which converges to a 1/\textit{f} slope. \par 
		
		Some caveats and corresponding justifications in the surface Alfv\'en wave setup are the following. We do not consider any effects that control plasma dynamics along the magnetic field of the background plasma. Thus we neglect gravity, heating and cooling terms, and thermal conduction. This is partly justified by the fact that surface Alfv\'en waves are mostly incompressible (therefore are weakly affected by thermal processes) and that we are only interested in dynamics perpendicular to the magnetic field. The lack of radial variations in the surface Alfv\'en wave setup is not such a drastic approximation as it may seem. At radial distances where the wave energy spectral slopes start showing change, at around 8 $R_\odot$, the density scale height in the model of \citet{2009ApJ...707.1659C} is over $10\ R_\odot$, which results in a rather shallow radial evolution of the Alfv\'en speed, evident in Fig.~\ref{fig2}. Radial expansion of the simulation box and of the background magnetic field is neglected. An effect of expansion is the shifting of perpendicular spatial scales to lower wavenumbers with increasing radial distance, as in the Alfv\'en wave setup. Effects related to dynamics along the field and expansion that are neglected here could have an impact on the ensuing linear perpendicular cascade by changing the density gradient across the field as a function of radial distance. This could accelerate or decelerate the energy cascade, depending on the background model used. For example, a varying degree of expansion could enhance Alfv\'en speed variations across magnetic field lines, while a homogeneous expansion could smooth them out. However, the dynamics presented here are fundamentally independent of such effects. The density maps used have a maximum density ratio of just above two, and local density ratios which are close to unity. This is done for both better numerical stability, and also due to the fact that large Alfv\'en speed gradients would take significantly longer to simulate, due to the increase in the largest Alfv\'en transit time. A higher density ratio would lead to a faster linear cascade. The simulation with a maximal density ratio of 10 confirms that the cascade speeds up and shows the depletion of the power at the driven, small wavenumbers, at which scales the wave energy tends towards a similar power law as at smaller scales. The slope of the power spectrum still remains determined by the underlying density map power spectrum, thus strengthening our conclusions. This simulation also shows that if the linear cascade rate is strong enough, the power law of the waves driven at the bottom of the corona is modified or lost through the linear cascade induced by the density inhomogeneity.   \par 
		
		Let us discuss next some underlying assumptions and implications of the presented results. A critical assumption in this study is the existence of density inhomogeneities with a well-defined power law. While the existence of such density inhomgeneities with various power laws is confirmed by numerous observational studies, as presented in the Introduction, their origin is less clear. While some studies on Alfv\'en wave turbulence show that the density perturbations act as a passive scalar, and pick up the energy spectrum of the velocity fluctuations, the real picture is probably much more complex, and involves the details of coronal heating and the coupled dynamics of the photosphere-chromosphere-corona system. Moreover, if the largest scales where the 1/\textit{f} spectrum is present are equivalent to an energy-containing range, then the underlying large-scale density fluctuations are probably less defined by the turbulence ensuing at the smaller scales, and more by driving at the solar surface. Thus, unlike in the inertial range, where the density fluctuation power law might be determined by the turbulent dynamics of velocity and magnetic fields, at large scales the density inhomogeneity power law determines the power law of the waves propagating through it, the other way around. 
		Another critical assumption is that we limit our study to linear processes. It is well known that waves propagating outwards from the Sun grow in (relative) amplitude as a consequence of energy conservation \citep[e.g.,][]{2001A&A...374L...9M}. As such, nonlinearities could become important and could alter the presented linear evolution in many ways. Therefore, it is important to highlight that our study is only relevant if the large scales where the 1/\textit{f} spectrum appears in the solar wind is dominated by linear evolution, as suggested by its WKB-like radial evolution.
		
		\begin{acknowledgments}
			
			TVD was supported by the European Research Council (ERC) under the European Union's Horizon 2020 research and innovation programme (grant agreement No 724326) and the C1 grant TRACEspace of Internal Funds KU Leuven. TVD has benefited from the funding of the FWO Vlaanderen through a Senior Research Project (G088021N)
			
		\end{acknowledgments}
		
		\bibliography{Biblio}{}

\begin{thebibliography}{}
\expandafter\ifx\csname natexlab\endcsname\relax\def\natexlab#1{#1}\fi
\providecommand{\url}[1]{\href{#1}{#1}}
\providecommand{\dodoi}[1]{doi:~\href{http://doi.org/#1}{\nolinkurl{#1}}}
\providecommand{\doeprint}[1]{\href{http://ascl.net/#1}{\nolinkurl{http://ascl.net/#1}}}
\providecommand{\doarXiv}[1]{\href{https://arxiv.org/abs/#1}{\nolinkurl{https://arxiv.org/abs/#1}}}

\bibitem[{{Bavassano} {et~al.}(1982){Bavassano}, {Dobrowolny}, {Mariani}, \&
  {Ness}}]{1982JGR....87.3617B}
{Bavassano}, B., {Dobrowolny}, M., {Mariani}, F., \& {Ness}, N.~F. 1982, \jgr,
  87, 3617, \dodoi{10.1029/JA087iA05p03617}

\bibitem[{{Belcher} \& {Davis}(1971)}]{1971JGR....76.3534B}
{Belcher}, J.~W., \& {Davis}, Jr., L. 1971, \jgr, 76, 3534,
  \dodoi{10.1029/JA076i016p03534}

\bibitem[{{Bruno} \& {Carbone}(2013)}]{2013LRSP...10....2B}
{Bruno}, R., \& {Carbone}, V. 2013, Living Reviews in Solar Physics, 10, 2,
  \dodoi{10.12942/lrsp-2013-2}

\bibitem[{{Bruno} {et~al.}(2004){Bruno}, {Carbone}, {Primavera}, {Malara},
  {Sorriso-Valvo}, {Bavassano}, \& {Veltri}}]{2004AnGeo..22.3751B}
{Bruno}, R., {Carbone}, V., {Primavera}, L., {et~al.} 2004, Annales
  Geophysicae, 22, 3751, \dodoi{10.5194/angeo-22-3751-2004}

\bibitem[{{Bruno} {et~al.}(2019){Bruno}, {Telloni}, {Sorriso-Valvo}, {Marino},
  {De Marco}, \& {D'Amicis}}]{2019A&A...627A..96B}
{Bruno}, R., {Telloni}, D., {Sorriso-Valvo}, L., {et~al.} 2019, \aap, 627, A96,
  \dodoi{10.1051/0004-6361/201935841}

\bibitem[{{Bruno} {et~al.}(2009){Bruno}, {Carbone}, {V{\"o}r{\"o}s},
  {D'Amicis}, {Bavassano}, {Cattaneo}, {Mura}, {Milillo}, {Orsini}, {Veltri},
  {Sorriso-Valvo}, {Zhang}, {Biernat}, {Rucker}, {Baumjohann},
  {Jankovi{\v{c}}ov{\'a}}, \& {Kov{\'a}cs}}]{2009EM&P..104..101B}
{Bruno}, R., {Carbone}, V., {V{\"o}r{\"o}s}, Z., {et~al.} 2009, Earth Moon and
  Planets, 104, 101, \dodoi{10.1007/s11038-008-9272-9}

\bibitem[{{Carley} {et~al.}(2021){Carley}, {Cecconi}, {Reid}, {Briand},
  {Sasikumar Raja}, {Masson}, {Dorovskyy}, {Tiburzi}, {Vilmer}, {Zucca},
  {Zarka}, {Tagger}, {Grie{\ss}meier}, {Corbel}, {Theureau}, {Loh}, \&
  {Girard}}]{2021ApJ...921....3C}
{Carley}, E.~P., {Cecconi}, B., {Reid}, H.~A., {et~al.} 2021, \apj, 921, 3,
  \dodoi{10.3847/1538-4357/ac1acd}

\bibitem[{{Chandran}(2018)}]{2018JPlPh..84a9006C}
{Chandran}, B. D.~G. 2018, Journal of Plasma Physics, 84, 905840106,
  \dodoi{10.1017/S0022377818000016}

\bibitem[{{Chandran} \& {Hollweg}(2009)}]{2009ApJ...707.1659C}
{Chandran}, B. D.~G., \& {Hollweg}, J.~V. 2009, \apj, 707, 1659,
  \dodoi{10.1088/0004-637X/707/2/1659}

\bibitem[{{Cranmer} \& {van Ballegooijen}(2005)}]{2005ApJS..156..265C}
{Cranmer}, S.~R., \& {van Ballegooijen}, A.~A. 2005, \apjs, 156, 265,
  \dodoi{10.1086/426507}

\bibitem[{{Cranmer} {et~al.}(2007){Cranmer}, {van Ballegooijen}, \&
  {Edgar}}]{2007ApJS..171..520C}
{Cranmer}, S.~R., {van Ballegooijen}, A.~A., \& {Edgar}, R.~J. 2007, \apjs,
  171, 520, \dodoi{10.1086/518001}

\bibitem[{{DeForest} {et~al.}(2018){DeForest}, {Howard}, {Velli}, {Viall}, \&
  {Vourlidas}}]{2018ApJ...862...18D}
{DeForest}, C.~E., {Howard}, R.~A., {Velli}, M., {Viall}, N., \& {Vourlidas},
  A. 2018, \apj, 862, 18, \dodoi{10.3847/1538-4357/aac8e3}

\bibitem[{{Denskat} \& {Neubauer}(1982)}]{1982JGR....87.2215D}
{Denskat}, K.~U., \& {Neubauer}, F.~M. 1982, \jgr, 87, 2215,
  \dodoi{10.1029/JA087iA04p02215}

\bibitem[{{D{\'\i}az-Su{\'a}rez} \& {Soler}(2021)}]{2021A&A...648A..22D}
{D{\'\i}az-Su{\'a}rez}, S., \& {Soler}, R. 2021, \aap, 648, A22,
  \dodoi{10.1051/0004-6361/202040161}

\bibitem[{{Dmitruk} \& {Matthaeus}(2007)}]{2007PhRvE..76c6305D}
{Dmitruk}, P., \& {Matthaeus}, W.~H. 2007, \pre, 76, 036305,
  \dodoi{10.1103/PhysRevE.76.036305}

\bibitem[{{Dmitruk} {et~al.}(2011){Dmitruk}, {Mininni}, {Pouquet}, {Servidio},
  \& {Matthaeus}}]{2011PhRvE..83f6318D}
{Dmitruk}, P., {Mininni}, P.~D., {Pouquet}, A., {Servidio}, S., \& {Matthaeus},
  W.~H. 2011, \pre, 83, 066318, \dodoi{10.1103/PhysRevE.83.066318}

\bibitem[{Dubey {et~al.}(2014)Dubey, Antypas, Calder, Daley, Fryxell,
  Gallagher, Lamb, Lee, Olson, Reid, Rich, Ricker, Riley, Rosner, Siegel,
  Taylor, Weide, Timmes, Vladimirova, \& ZuHone}]{Dubey2013}
Dubey, A., Antypas, K., Calder, A.~C., {et~al.} 2014, International Journal of
  High Performance Computing Applications, 28, 225,
  \dodoi{http://dx.doi.org/10.1177/1094342013505656}

\bibitem[{{Els\"{a}sser}(1950)}]{1950PhRv...79..183E}
{Els\"{a}sser}, W.~M. 1950, Physical Review, 79, 183,
  \dodoi{10.1103/PhysRev.79.183}

\bibitem[{{Federrath} {et~al.}(2010){Federrath}, {Roman-Duval}, {Klessen},
  {Schmidt}, \& {Mac Low}}]{2010A&A...512A..81F}
{Federrath}, C., {Roman-Duval}, J., {Klessen}, R.~S., {Schmidt}, W., \& {Mac
  Low}, M.~M. 2010, \aap, 512, A81, \dodoi{10.1051/0004-6361/200912437}

\bibitem[{{Gogoberidze} \& {Voitenko}(2016)}]{2016Ap&SS.361..364G}
{Gogoberidze}, G., \& {Voitenko}, Y.~M. 2016, \apss, 361, 364,
  \dodoi{10.1007/s10509-016-2950-6}

\bibitem[{{Goossens} {et~al.}(2002){Goossens}, {de Groof}, \&
  {Andries}}]{2002ESASP.505..137G}
{Goossens}, M., {de Groof}, A., \& {Andries}, J. 2002, in ESA Special
  Publication, Vol. 505, SOLMAG 2002. Proceedings of the Magnetic Coupling of
  the Solar Atmosphere Euroconference, ed. H.~{Sawaya-Lacoste}, 137--144

\bibitem[{{Goossens} {et~al.}(2009){Goossens}, {Terradas}, {Andries},
  {Arregui}, \& {Ballester}}]{2009A&A...503..213G}
{Goossens}, M., {Terradas}, J., {Andries}, J., {Arregui}, I., \& {Ballester},
  J.~L. 2009, \aap, 503, 213, \dodoi{10.1051/0004-6361/200912399}

\bibitem[{{Goossens} {et~al.}(2019){Goossens}, {Arregui}, \& {Van
  Doorsselaere}}]{2019FrASS...6...20G}
{Goossens}, M.~L., {Arregui}, I., \& {Van Doorsselaere}, T. 2019, Frontiers in
  Astronomy and Space Sciences, 6, 20, \dodoi{10.3389/fspas.2019.00020}

\bibitem[{Harris {et~al.}(2020)Harris, Millman, van~der Walt, Gommers,
  Virtanen, Cournapeau, Wieser, Taylor, Berg, Smith, Kern, Picus, Hoyer, van
  Kerkwijk, Brett, Haldane, del R{\'{i}}o, Wiebe, Peterson,
  G{\'{e}}rard-Marchant, Sheppard, Reddy, Weckesser, Abbasi, Gohlke, \&
  Oliphant}]{harris2020array}
Harris, C.~R., Millman, K.~J., van~der Walt, S.~J., {et~al.} 2020, Nature, 585,
  357, \dodoi{10.1038/s41586-020-2649-2}

\bibitem[{{Heinemann} \& {Olbert}(1980)}]{1980JGR....85.1311H}
{Heinemann}, M., \& {Olbert}, S. 1980, \jgr, 85, 1311,
  \dodoi{10.1029/JA085iA03p01311}

\bibitem[{{Heyvaerts} \& {Priest}(1983)}]{1983A&A...117..220H}
{Heyvaerts}, J., \& {Priest}, E.~R. 1983, \aap, 117, 220

\bibitem[{{Hnat} {et~al.}(2005){Hnat}, {Chapman}, \&
  {Rowlands}}]{2005PhRvL..94t4502H}
{Hnat}, B., {Chapman}, S.~C., \& {Rowlands}, G. 2005, \prl, 94, 204502,
  \dodoi{10.1103/PhysRevLett.94.204502}

\bibitem[{{Hollweg}(1973)}]{1973JGR....78.3643H}
{Hollweg}, J.~V. 1973, \jgr, 78, 3643, \dodoi{10.1029/JA078i019p03643}

\bibitem[{{Hollweg}(1990)}]{1990JGR....9514873H}
---. 1990, \jgr, 95, 14873, \dodoi{10.1029/JA095iA09p14873}

\bibitem[{{Ionson}(1978)}]{1978ApJ...226..650I}
{Ionson}, J.~A. 1978, \apj, 226, 650, \dodoi{10.1086/156648}

\bibitem[{{Kasper} {et~al.}(2021){Kasper}, {Klein}, {Lichko}, {Huang}, {Chen},
  {Badman}, {Bonnell}, {Whittlesey}, {Livi}, {Larson}, {Pulupa}, {Rahmati},
  {Stansby}, {Korreck}, {Stevens}, {Case}, {Bale}, {Maksimovic}, {Moncuquet},
  {Goetz}, {Halekas}, {Malaspina}, {Raouafi}, {Szabo}, {MacDowall}, {Velli},
  {Dudok de Wit}, \& {Zank}}]{2021PhRvL.127y5101K}
{Kasper}, J.~C., {Klein}, K.~G., {Lichko}, E., {et~al.} 2021, \prl, 127,
  255101, \dodoi{10.1103/PhysRevLett.127.255101}

\bibitem[{Lithwick \& Goldreich(2001)}]{0004-637X-562-1-279}
Lithwick, Y., \& Goldreich, P. 2001, The Astrophysical Journal, 562, 279.
\newblock \url{http://stacks.iop.org/0004-637X/562/i=1/a=279}

\bibitem[{{Magyar} \& {Nakariakov}(2021)}]{2021ApJ...907...55M}
{Magyar}, N., \& {Nakariakov}, V.~M. 2021, \apj, 907, 55,
  \dodoi{10.3847/1538-4357/abd02f}

\bibitem[{{Malara}(2013)}]{2013A&A...549A..54M}
{Malara}, F. 2013, \aap, 549, A54, \dodoi{10.1051/0004-6361/201219307}

\bibitem[{{Marsch} \& {Tu}(1990)}]{1990JGR....9511945M}
{Marsch}, E., \& {Tu}, C.~Y. 1990, \jgr, 95, 11945,
  \dodoi{10.1029/JA095iA08p11945}

\bibitem[{{Matteini} {et~al.}(2018){Matteini}, {Stansby}, {Horbury}, \&
  {Chen}}]{2018ApJ...869L..32M}
{Matteini}, L., {Stansby}, D., {Horbury}, T.~S., \& {Chen}, C.~H.~K. 2018,
  \apjl, 869, L32, \dodoi{10.3847/2041-8213/aaf573}

\bibitem[{{Matthaeus} {et~al.}(2007){Matthaeus}, {Breech}, {Dmitruk},
  {Bemporad}, {Poletto}, {Velli}, \& {Romoli}}]{2007ApJ...657L.121M}
{Matthaeus}, W.~H., {Breech}, B., {Dmitruk}, P., {et~al.} 2007, \apjl, 657,
  L121, \dodoi{10.1086/513075}

\bibitem[{{Matthaeus} \& {Goldstein}(1986)}]{1986PhRvL..57..495M}
{Matthaeus}, W.~H., \& {Goldstein}, M.~L. 1986, \prl, 57, 495,
  \dodoi{10.1103/PhysRevLett.57.495}

\bibitem[{{Matthaeus} {et~al.}(1999){Matthaeus}, {Zank}, {Oughton}, {Mullan},
  \& {Dmitruk}}]{1999ApJ...523L..93M}
{Matthaeus}, W.~H., {Zank}, G.~P., {Oughton}, S., {Mullan}, D.~J., \&
  {Dmitruk}, P. 1999, \apjl, 523, L93, \dodoi{10.1086/312259}

\bibitem[{{Moncuquet} {et~al.}(2020){Moncuquet}, {Meyer-Vernet}, {Issautier},
  {Pulupa}, {Bonnell}, {Bale}, {Dudok de Wit}, {Goetz}, {Griton}, {Harvey},
  {MacDowall}, {Maksimovic}, \& {Malaspina}}]{2020ApJS..246...44M}
{Moncuquet}, M., {Meyer-Vernet}, N., {Issautier}, K., {et~al.} 2020, \apjs,
  246, 44, \dodoi{10.3847/1538-4365/ab5a84}

\bibitem[{{Moran}(2001)}]{2001A&A...374L...9M}
{Moran}, T.~G. 2001, \aap, 374, L9, \dodoi{10.1051/0004-6361:20010643}

\bibitem[{{Morton} {et~al.}(2015){Morton}, {Tomczyk}, \&
  {Pinto}}]{2015NatCo...6E7813M}
{Morton}, R.~J., {Tomczyk}, S., \& {Pinto}, R. 2015, Nature Communications, 6,
  7813, \dodoi{10.1038/ncomms8813}

\bibitem[{{Parker}(1991)}]{1991ApJ...376..355P}
{Parker}, E.~N. 1991, \apj, 376, 355, \dodoi{10.1086/170285}

\bibitem[{{Pascoe} {et~al.}(2010){Pascoe}, {Wright}, \& {De
  Moortel}}]{2010ApJ...711..990P}
{Pascoe}, D.~J., {Wright}, A.~N., \& {De Moortel}, I. 2010, \apj, 711, 990,
  \dodoi{10.1088/0004-637X/711/2/990}

\bibitem[{{Perez} \& {Chandran}(2013)}]{2013ApJ...776..124P}
{Perez}, J.~C., \& {Chandran}, B.~D.~G. 2013, \apj, 776, 124,
  \dodoi{10.1088/0004-637X/776/2/124}

\bibitem[{{Rast}(2003)}]{2003ApJ...597.1200R}
{Rast}, M.~P. 2003, \apj, 597, 1200, \dodoi{10.1086/381221}

\bibitem[{{Raymond} {et~al.}(2014){Raymond}, {McCauley}, {Cranmer}, \&
  {Downs}}]{2014ApJ...788..152R}
{Raymond}, J.~C., {McCauley}, P.~I., {Cranmer}, S.~R., \& {Downs}, C. 2014,
  \apj, 788, 152, \dodoi{10.1088/0004-637X/788/2/152}

\bibitem[{{Shoda} \& {Yokoyama}(2018)}]{2018ApJ...859L..17S}
{Shoda}, M., \& {Yokoyama}, T. 2018, \apjl, 859, L17,
  \dodoi{10.3847/2041-8213/aac50c}

\bibitem[{{Soler} \& {Terradas}(2015)}]{2015ApJ...803...43S}
{Soler}, R., \& {Terradas}, J. 2015, \apj, 803, 43,
  \dodoi{10.1088/0004-637X/803/1/43}

\bibitem[{{Tomczyk} {et~al.}(2007){Tomczyk}, {McIntosh}, {Keil}, {Judge},
  {Schad}, {Seeley}, \& {Edmondson}}]{2007Sci...317.1192T}
{Tomczyk}, S., {McIntosh}, S.~W., {Keil}, S.~L., {et~al.} 2007, Science, 317,
  1192, \dodoi{10.1126/science.1143304}

\bibitem[{{Van Doorsselaere} {et~al.}(2008){Van Doorsselaere}, {Nakariakov}, \&
  {Verwichte}}]{2008ApJ...676L..73V}
{Van Doorsselaere}, T., {Nakariakov}, V.~M., \& {Verwichte}, E. 2008, \apjl,
  676, L73, \dodoi{10.1086/587029}

\bibitem[{{Velli} {et~al.}(1989){Velli}, {Grappin}, \&
  {Mangeney}}]{1989PhRvL..63.1807V}
{Velli}, M., {Grappin}, R., \& {Mangeney}, A. 1989, Physical Review Letters,
  63, 1807, \dodoi{10.1103/PhysRevLett.63.1807}

\bibitem[{{Verdini} {et~al.}(2012){Verdini}, {Grappin}, {Pinto}, \&
  {Velli}}]{2012ApJ...750L..33V}
{Verdini}, A., {Grappin}, R., {Pinto}, R., \& {Velli}, M. 2012, \apjl, 750,
  L33, \dodoi{10.1088/2041-8205/750/2/L33}

\bibitem[{{Verdini} {et~al.}(2009){Verdini}, {Velli}, \&
  {Buchlin}}]{2009ApJ...700L..39V}
{Verdini}, A., {Velli}, M., \& {Buchlin}, E. 2009, \apjl, 700, L39,
  \dodoi{10.1088/0004-637X/700/1/L39}

\bibitem[{{{\v{S}}afr{\'a}nkov{\'a}} {et~al.}(2015){{\v{S}}afr{\'a}nkov{\'a}},
  {N{\v{e}}me{\v{c}}ek}, {N{\v{e}}mec}, {P{\v{r}}ech}, {Pit{\v{n}}a}, {Chen},
  \& {Zastenker}}]{2015ApJ...803..107S}
{{\v{S}}afr{\'a}nkov{\'a}}, J., {N{\v{e}}me{\v{c}}ek}, Z., {N{\v{e}}mec}, F.,
  {et~al.} 2015, \apj, 803, 107, \dodoi{10.1088/0004-637X/803/2/107}

\bibitem[{{Woo} \& {Armstrong}(1979)}]{1979JGR....84.7288W}
{Woo}, R., \& {Armstrong}, J.~W. 1979, \jgr, 84, 7288,
  \dodoi{10.1029/JA084iA12p07288}

\end{thebibliography}
		\bibliographystyle{aasjournal}
		
	\end{document}